\documentclass[prb,showpacs,preprintnumbers,superscriptaddress,11pt]{revtex4}
\newif\ifpdf
\ifx\pdfoutput\undefined
   \pdffalse
\else
   \pdfoutput=1
   \pdftrue
\fi

\ifpdf
 \usepackage[pdftex]{graphicx}
\else
 \usepackage[dvips]{graphicx}
\fi

\usepackage{dcolumn}
\usepackage{bm}
\usepackage{amssymb}
\usepackage{amsmath}
\usepackage{amsfonts}
\usepackage{latexsym}
\usepackage{hhline}
\usepackage{theorem}

\numberwithin{equation}{section}
\newcommand{\eqr}[1]{Eq.(\ref{#1})}
\newcommand{\figr}[1]{Figure \ref{#1}}
\newtheorem{theorem}{Theorem}[section]

\theoremstyle{plain}{\theorembodyfont{\rmfamily}%
}
\theoremstyle{plain}{\theorembodyfont{\rmfamily}%
\newtheorem{algorithm}[theorem]{Algorithm}}
\theoremstyle{plain}{\theorembodyfont{\rmfamily}%
}
\theoremstyle{plain}{\theorembodyfont{\rmfamily}%
}

\newcommand{\Jbb}{\mathbb{J}}
\newcommand{\Rbb}{\mathbb{R}}

\newcommand{\Yhat}{{\widehat{Y}}}
\newcommand{\Shat}{{\widehat{S}}}

\newcommand{\gbar}{{\overline{g}}}

\renewcommand{\equiv}{:=}

\newcommand{\ucmin}[2]{\underset{#2}{\mbox{minimize}}~#1}

\newcommand{\und}{\quad\mbox{ and }\quad}
\newcommand{\where}{\quad\mbox{ where }\quad}

\DeclareMathOperator{\prox}{prox}

\DeclareMathOperator{\argmin}{argmin\,}
\DeclareMathOperator{\dist}{dist \;}

\begin{document}

\preprint{APS/123-QED}

\title{Robust Mixing for Ab-Initio Quantum Mechanical Calculations}
\author{L. D. Marks}
\affiliation{Department of Materials Science and Engineering, 
Northwestern University, Evanston, IL 60201.}\;
\author{D. R. Luke}
\email{rluke@math.udel.edu}
\affiliation{Department of Mathematical Sciences, University of Delaware, Newark DE
  19716-2553, USA.}\;

\date{version 0.98---January 10, 2008}

\begin{abstract}
We study the general problem of 
mixing for ab-initio quantum-mechanical 
problems.  Guided by general mathematical principles and the underlying physics, we propose
a multisecant form of Broyden's second method for solving the self-consistent field
equations of Kohn-Sham density functional theory.  
The algorithm is robust, requires relatively little fine-tuning and appears to
outperform the current state of the art, converging for cases that defeat 
many other methods.  We compare our technique to 
the conventional methods for problems ranging from simple to nearly 
pathological. 
\end{abstract}


\maketitle

\section{\label{s:intro} Introduction}
We consider the problem of determining the electron density $\rho$ that satisfies
the self-consistent field equations according to the Kohn-Sham density functional 
theory \cite{KohnSham65, Mermin65}:  
\begin{subequations}\label{e:scf}
\begin{equation}\label{e:scf1}
(H_0+ V_\rho)\phi_i = \epsilon_i \phi_i
\end{equation}
\begin{equation}\label{e:scf2}
\rho(x) = \sum_i \left(1+ e^{\beta(\epsilon_i - \mu)} \right)^{-1}|\phi_i(x)|^2.
\end{equation}
\end{subequations}
Here $H_0$ is the single-particle noninteracting Hamiltonian and $V_\rho$ 
is an effective potential parameterized by the particle density $\rho$.  The 
constant $\beta$ is $1/kT$ where $k$ is Boltzmann's constant and $T$ is temperature. 
The term $(1+ e^{\beta(\epsilon_i - \mu)})^{-1} $ is the Fermi-Dirac occupation
and the 
constant $\mu$ is determined by $\int \rho(x) dx  = N$ for an $N$-body problem.    
Following \cite{Prodan05} we let $H_\rho\equiv H_0 + \lambda V_\rho$ 
\footnote{``$\equiv$'' distinguishes a {\em definition} from an equation.} denote 
the Kohn-Sham Hamiltonian and reformulate the above system of equations 
as a nonlinear fixed point problem: 
find $\rho$ such that 
\begin{equation}\label{e:fix}
F(\rho)(x) \equiv \left(1+ e^{\beta(H_\rho - \mu)} \right)^{-1}(x,x)=\rho(x)
\end{equation}
where $\mu$ is the unique solution to $N=trace(\left(1+ e^{\beta(H_\rho - \mu)} \right)^{-1})$.  
We refer to the operator $F$ above as the self-consistent field (SCF) operator.  
We will not be concerned with the details of the SCF operator or its 
approximations since these tend to be specific to the application.  Also, we will work with 
the discretized version of the SCF operator, which we will call the SCF {\em mapping} since
it is a real vector-valued mapping of the discretized density.  Throughout this work, 
however, we will point to instances where the form of this mapping can cause
problems for numerical procedures for solving \eqr{e:fix}.  

Numerical algorithms for solving \eqr{e:fix} abound --
the representative examples we focus on here are 
\cite{Wien2k, CrittinBierlaire03, Johnson88,Kawata98,Pulay80, Srivastava,VanderbiltLouie84, Singh86}.  
These are iterative procedures
and the process of determining the desired density $\rho$ from previous estimates 
has come to be known as ``mixing'' in the physical literature.  For ab-initio methods 
there is frequently a user-provided mixing term which, if it is 
improperly chosen, will lead to divergence of the iterations. In many cases the user has to 
learn by failure what is the correct value to use, expending a fair amount of computer 
resources in the process.  We will show that many of the methods found in the physical 
literature have counterparts in the mathematical literature thus providing a more
systematic approach to the choice of algorithm parameters.    
The goal of this work is the development of a method that does not require expert 
user input, is fast, and can handle 
many of the more complicated and poorly convergent problems such as metallic surfaces or 
heterostructures that can defeat a novice and sometimes an expert.

In the next section we provide detailed background both 
to the mathematical literature on methods 
for solving non-linear equations, as well the physical literature on mixing.
There are two major ``branches'' of numerical techniques distinguished by the space 
in which they are derived.  However we show that all of the different methods are generated by 
solving an optimization problem of the same form. In Section \ref{s:new stuff} we detail 
our proposed algorithm for solving \eqr{e:fix}.  
We show in Section \ref{s:new stuff} that most algorithms are predicated upon fixed point 
mappings with a great deal of regularity, in particular monotone mappings.  Motivated by 
the prospect that most systems of physical interest do not lead to convex SCF mappings, the 
principal insight that yields our improvements
is to treat the prior steps as random samples 
in a higher-dimensional space rather than deterministic steps in a path to a fixed point. 
In effect, the Kohn-Sham Hamiltonian evaluated at an electron density far
from equilibrium can vary in a semirandom and perhaps chaotic fashion.
This viewpoint leads to a natural interpretation of the updates in terms of 
predicted and unpredicted components.  Controls on the algorithm are 
exerted through parameter choices that affect primarily the unpredicted 
step.   The rest of the section involves technical considerations for controlling step sizes
and safeguarding against instabilities.
We introduce the idea of using proximal mappings \cite{Moreau62} to 
account for problems associated with near-linear dependencies and, at the same time,
to safeguard against uncertainties in the model  
analogous to a classic Wiener filter.  This regularization has the 
advantage that it also acts similarly to a classic trust region technique in numerical optimization. 
In Section \ref{s:results} we present 
numerical results for both very easy problems as well as semi-pathological cases. 
The new approach outperforms existing algorithms in 
most cases, and does significantly better with poorly constructed Kohn-Sham mappings. The algorithm is 
also relatively insensitive to user input. We conclude with a discussion of some of the open issues.

\section{\label{s:background} Iterative methods for solving nonlinear equations}
We wish to determine the electron density $\rho_*$ 
with fixed atom locations.   The density is a real-valued vector with $k$ elements. With an 
estimated density $\rho_{n}$ at the  $n$-th step of an iterative procedure for 
determining $\rho_*$, we check whether our estimate satisfies the ab-initio self-consistent 
field (SCF) equations given by \eqr{e:fix}.  
Evaluation of the SCF mapping returns a modified density $\rho_{n}'\equiv F(\rho_n)$, 
another real-valued vector with $k$ elements.
The density we seek is a fixed point of $F$, i.e., we solve the system of non-linear equations
\begin{equation}\label{e:nleq}
	F(\rho_*)-\rho_*=0.										
\end{equation}
This suggests the usual Newton algorithm as a possible numerical solution strategy.  

\subsection{\label{s:Newton} Newton's Method}
Expressing the mapping in \eqr{e:nleq}  by its Taylor series expansion centered on a fixed 
point $\rho_*$  yields
\begin{equation}\label{e:Taylor}
	 F(\rho) - \rho  = (J(\rho_*) - I)(\rho - \rho_*) + O(|\rho-\rho_*|^2)     \end{equation}

where $J(\rho_*)$ is the Jacobian of $F$ (supposing this is well-defined) at $\rho_* $ and 
$I$ is the identity mapping.   The Jacobian $J$ is a $k$-by-$k$ real-valued matrix.  Given a 
point  $\rho_{n}$, Newton's method generates the next point $\rho_{n+1}$ by
\begin{equation}\label{e:rho_n+1}
	 \rho_{n+1} =  \rho_{n} - (J(\rho_{n})-I)^{-1}(F(\rho_{n})-\rho_{n}).			
\end{equation}

Under standard assumptions, this iteration can be shown to converge quadratically in a neighborhood 
of a local solution \cite{BoydVan03}.  The computational cost of calculating the Jacobian and 
inverting is, however, prohibitive  for high-dimensional problems such as density functional calculations.  
 
\subsection{Matrix Secant Methods}
A classical approach to avoid computing and inverting the Jacobian is via solutions to the 
matrix secant equation:  $(J(\rho_{n}) - I) \approx B_{n}$ 
where 
\begin{equation}\label{e:mat seq}
	B_{n}(\rho_{n} - \rho_{n-1}) =
\Big((F(\rho_{n}) - \rho_{n})-(F(\rho_{n-1}) - \rho_{n-1})\Big)				
\end{equation}
Introducing new variables, a conventional reduction, this is compactly represented as 
\begin{equation}\label{e:forward}
	 B_{n} s_{n-1} = y_{n-1} 
\end{equation}
or
\begin{equation}\label{e:backward}
	  H_{n} y_{n-1} = s_{n-1} 		 
\end{equation}
where  $H_{n}=B_{n}^{-1}$ and 
\begin{equation}\label{e:sn yn}
s_{n-1} = \rho_{n} - \rho_{n-1} \und  y_{n-1} = (F(\rho_{n})  - \rho_{n}) - (F(\rho_{n-1}) - \rho_{n-1}). 	
\end{equation}

The next density $\rho_{n+1}$ is then generated either by the recursion
\begin{equation}\label{e:rho_n+1 forward}
\rho_{n+1} =  \rho_{n} -  B_{n}^{-1}(F(\rho_{n}) - \rho_{n})
\end{equation}
where $B_{n}$ satisfies \eqr{e:forward}, or by 
\begin{equation}\label{e:rho_n+1 backward}
\rho_{n+1} =  \rho_{n} -  H_{n}(F(\rho_{n}) - \rho_{n})	
\end{equation}
where $H_{n}$ satisfies \eqr{e:backward}.  Sequences generated by \eqr{e:rho_n+1 forward} or 
\eqr{e:rho_n+1 backward} are known as quasi-Newton methods.   The unknowns in 
\eqr{e:forward} and \eqr{e:backward} are the matrices $B_{n}$ and $H_{n}$, hence these are 
systems of $k$ linear equations in $k^2$ unknowns.  There are infinitely many solutions 
to the matrix secant equation 
and each different solution leads to a different numerical method for finding 
the fixed point of $F$.   A common approach in the 
literature on density functional calculations, due to Srivastava
\cite{Srivastava}, is based on the Broyden rank one matrix updates \cite{Broyden65}.  
Other common updates in the optimization literature are the symmetric rank one  
(SR1) and the BFGS updates \cite{DennisSchnabel96}.  
Our focus in this study is on improvements in the context of Broyden updates, however the basic 
principles outlined here extend more generally to other matrix secant methods.

\subsection{Rank One Updates}
A new matrix $B_{n+1}$  is obtained by updating in some fashion $B_{n}$  using the new data 
pair $(s_{n},y_{n})$ combined with the prior information $(s_{0},y_{0}), (s_{1},y_{1}),..., (s_{n-1},y_{n-1})$ 
subject to the constraint that  $B_{n+1}$  satisfy \eqr{e:forward}.   In his original paper Broyden 
\cite{Broyden65} looked at two approaches. The first, normally referred to as Broyden's ``good'' 
method (GB), finds the nearest matrix to $B_{n}$  with respect to the Frobenius norm 
(i.e. the square root of the sum of squares of 
the matrix entries) that solves 
\eqr{e:forward} (this property, due to Dennis and Mor\'e \cite{DennisMore77} is slightly different 
than Broyden's original interpretation).  This is given explicitly by 
\begin{equation}\label{e:GB}
	B_{n+1}=B_{n}+\frac{(y_{n}-\beta_nB_{n}s_{n})s_{n}^{T}}{\beta_n||s_{n}||^2}.
\end{equation}
Here and throughout this work the norm $||s||=\sqrt{s^Ts}$ is the Euclidean 
norm and a vector (understood to be a {\em column} vector) or matrix 
raised to the power $^T$ indicates the transpose. 
The scalar $\beta_n>0$ is a step size parameter which has been retained for formal 
consistency although it is normally taken to be unity.
This method has a simple recursion for its inverse (needed in \eqr{e:rho_n+1 forward}):
\begin{equation}\label{e:GBinv}
	B_{n+1}^{-1}= B_{n}^{-1}+ 
\frac{\left(\beta_ns_{n} - B_{n}^{-1}y_{n}\right)s_{n}^{T}B_{n}^{-1}}%
{(s_{n}^{T}B_{n}^{-1}y_{n})}.
\end{equation} 
Broyden's first method is shown in \cite[Theorem 5.2]{DennisMore77} to converge locally superlinearly 
under the standard assumptions that the Jacobian exists, is nonsingular, and Lipschitz continuous at 
the solution.  

The second method proposed by Broyden has the attractive feature that it is a recursion on the inverse 
Jacobian $H_{n}$  and is given by 
\begin{equation}\label{e:BB}
	H_{n+1} = H_{n} + 
\frac{\left(\beta_ns_{n} - H_{n}y_{n}\right)y_{n}^{T}}{||y_{n}||^2}.
\end{equation} 
Note that our sign convention is different to the usual sign in the literature 
where  $H_{n+1}=-B_{n+1}^{-1}$.  Update \eqr{e:BB} was not recommended by Broyden and subsequently 
became known as Broyden's ``bad'' method (BB)  primarily, in our opinion, because the test problems to 
which the methods were applied favored \eqr{e:GB}.  Analogously to GB, the BB update finds the 
nearest matrix to $H_{n}$  with respect to the Frobenius norm that solves \eqr{e:backward}.  

Generalizing the two methods suggested by Broyden one could consider
updates of the form
\begin{equation}\label{e:Barnes}
B_{n+1} = B_{n} + 
\frac{\left(y_{n} -B_{n}s_{n}\right)v_{n}^{T}}{v_{n}^Ts_{n}}.
\end{equation} 
Barnes \cite{Barnes65} and Gay and Schnabel \cite{GaySchnabel78} proposed an update of this form 
where $v_n$ is the projection of $s_n$ onto the orthogonal 
complement of $[s_{n-m}, s_{n-m+1}, \dots, s_n]$ ($0\leq m< n$).  This strategy, known as 
{\em projected updates}, will be discussed in more detail in the next subsection.  

A variation of \eqr{e:BB} for the inverse recursion $H_n$ was proposed 
by Martinez and Zambaldi \cite{Martinez92c} that takes the form
\begin{equation}\label{e:Martinez}
H_{n+1} = H_{n} + 
\frac{\left(s_{n} -H_{n}y_{n}\right)e_{n,j}^{T}}{e_{n,j}^Ty_{n}}
\end{equation} 
where $e_{n,j}$ is the $j$'th column of the identity matrix, chosen so that 
$\left|e_{n,j}^Ty_n\right| = \|y_n\|_{\infty}$ where $\|y_n\|_{\infty}$ is the 
component of $y_n$ with largest absolute value.  The matrix update then differs
from the previous update only in the $j$'th column. Numerical experience 
with this update is very favorable \cite{Spedicato97, LucksanVlcek98,
Martinez00}.

Recall that the GB and BB updates are the matrices nearest to the previous matrices
with repsect to the Frobenius norm that satisfy the matrix secant equation.  The main 
difference between the GB and BB methods  is the space in which the ``nearest'' 
criterion is applied \cite{IPTodd88}.  For GB the criterion is applied in the domain 
of the mapping, while BB is applied in the range, where the domain of the mapping is the 
space of the density differences $s_n$ and the range is the space of the residual differences $y_n$.  
We see from \eqr{e:rho_n+1 forward} 
that an ill-conditioned matrix update $B_{n}$ will lead to a large and numerically unstable 
estimation of the step $s_{n}$ since, to compute this one needs to invert $B_{n}$.   On the other hand, 
from \eqr{e:rho_n+1 backward} it is clear that a least change criterion in the space 
of the residual differences $y_n$ will lead to
 smaller steps that could slow progress for well-conditioned problems. In other words, the BB algorithm 
is inherently more conservative than the GB algorithm.  The advantage of this property is not apparent 
until one is faced with ill-conditioned or ill-posed problems -- see Section \ref{s:new stuff}
for a discussion in the context of
the physics of a DFT problem.   In this respect, the distinction between 
Broyden's first and second methods is analogous to the distinction between backward, or implicit,  
and forward techniques for numerical solutions to stiff differential equations.  
We formulate these statements more 
precisely below.

The recursions \eqr{e:GB}-(\ref{e:Barnes}) are all
one-step recursions in terms of the most recent Jacobian estimate.  Most practical
implementations make use of multi-step recursions on the inverse Jacobian 
that allow one to avoid explicitly forming the matrix.  In particular, the recursion for 
\eqr{e:GBinv} with $\beta_j=1$ ($j=0,1,2,\dots$), can be written
as \cite[Theorem 6.2]{ByrdNocedalSchnabel94}
\begin{equation}\label{e:GBinv0}
B_{n+1}^{-1} = B_{0}^{-1} - \left(B_{0}^{-1}Y_n - S_n \right)
\left( L_n+S_n^TB_{0}^{-1}Y_n \right)^{-1}S_n^TB_{0}^{-1}
\end{equation}
where 
\begin{equation}\label{SnYn}
S_n\equiv \left[s_0, s_1, s_2, \dots,s_n \right],\qquad 
Y_n\equiv \left[y_0, y_1, y_2, \dots,y_n \right]\qquad (\mbox{$k$-by-$(n+1)$ matrices})
\end{equation}
and 
\[
(L_n)_{i,j}\equiv\begin{cases}-s_{i-1}^Ts_{j-1}&\mbox{ if }~ i>j\\
0& \mbox{else}
\end{cases}.
\]
For the multi-step recursion of $H_{n+1}$ an elementary calculation 
yields the following recursion for \eqr{e:BB}:
\begin{equation}\label{e:Hrec}
H_{n+1} = H_{0}\prod_{j=0}^n W_{j} - \sum_{j=0}^n\left( Z_{j}\prod_{i=j+1}^nW_i\right).
\end{equation}
where the products ascend from left to right with the empty product defined as $1$, and
\[
W_j\equiv I - \frac{y_jy_j^T}{\|y_j\|^2}
\und
Z_j\equiv \beta_j \frac{s_jy_j^T}{\|y_j\|^2}.
\]
We prefer this recursion over the recursion proposed
by Srivastava \cite{Srivastava} because, as we shall see in the following sections,  
we gain valuable insight into the geometry of the operations for the same storage requirements.
Srivastava's formulation was implimented for LAPW code by \cite{Singh86}, although extensive 
numerical experience in the Wien2k code indicates that $H_0$ needs to be adjusted 
dynamically, for reasons that will be clearer later. 

In the recursions \eqr{e:GBinv0} and \eqr{e:Hrec} the initial matrix, $B_0$ and $H_0$ respectively,
is crucial.  Several authors have studied optimal initializations 
\cite{Davidon59, Davidon75, Oren, Phua, Schnabel78, IPTodd88}.  
We explore scalings in greater detail in Subsection \ref{s:step}. 
A few papers have also explored strategies for combining the updates \eqr{e:GBinv0}
and \eqr{e:Hrec} in order to take advantage of the strengths of each.  IP and Todd 
\cite{IPTodd88} derive an update that is a 
convex combination of GB and BB with weighting determined explicitly so as to yield an 
optimally conditioned matrix.   Their method is locally superlinearly 
convergent \cite[Corollary 11]{IPTodd88} under standard assumptions and the additional conditions
that the initial approximate Jacobian, $B_0,$ is close to the true Jacobian 
at the solution and that the sequence of matrix updates and their inverses are uniformly bounded.
Martinez \cite{Martinez82} switches between GB and BB based on a simple criterion that 
estimates which method will give the best performance at a given step.  
Numerical experiments showed 
this to be a promising approach, but rates of convergence were not pursued.  

\subsection{\label{s:prox} Multisecant Methods}
In the context of path following algorithms, it is natural to look only at local information in 
order to construct a step direction and length.  To generate the $n+1$-th Jacobian approximation 
the methods described above satisfy the matrix secant equation \eqr{e:forward} or \eqr{e:backward} 
for the current step $s_n$ and residual difference $y_n$.    However, for highly nonlinear problems, 
where iterates may belong to domains of attraction of different solutions, it is perhaps more 
appropriate to view the previous 
data as samples of an unknown process in a high dimensional space.  In this context, updating 
the Jacobian based only on the most recent sample and ignoring  
the other sample points imposes a bias.  Invoking some of the terminology used for stochastic 
optimization, searching for the nearest matrix that satisfies the matrix secant equation only for 
the most recent sample point is a greedy strategy without recourse.  

Multisecant techniques put the previous data on more equal footing with the most 
recent steps; that is, rather than satisfying the matrix secant equation for only the most 
recent step one satisfies {\em all} matrix secant equations {\em simultaneously}:
\begin{equation}\label{e:multisecant}
Y_n = B_{n}S_n, 
\end{equation}
where $S_n=[s_{1,n},s_{2,n},\dots,s_{m,n}]$ and 
$Y_n=[y_{1,n},y_{2,n},\dots,y_{m,n}]$ are $k$-by-$m$ ($m\leq \min\{n,k\}$) matrices 
whose columns are previous steps and residual differences respectively.   
Multisecant techniques have been studied by several authors 
in the mathematical literature over the last $48$ years 
\cite{Wolfe59, Barnes65, OrtegaRheinboldt70, GraggStewart76, GaySchnabel78, Martinez79, 
Schnabel83, 
ByrdNocedalSchnabel94, DennisSchnabel96, FordMoghrabi97}.
A few authors in the physical sciences 
\cite{Pulay80,VanderbiltLouie84, Johnson88, Kawata98, CrittinBierlaire03} independently 
derived updates that, we will show, are simple relaxations of more conventional multisecant methods.  

For our 
application the dimension 
of the problem, $k$,  is much larger than the number of iterations, $n$, so we will 
simply consider $m$ matrix secant equations with $m<n$.  
We do not as yet specify the composition of 
$S_n$ and $Y_n$ since
there are many options.  The construction given in \eqr{e:sn yn} 
is conventional for matrix secant methods;  an alternative construction
centers the steps on the initial point $\rho_0$ rather than the previous step as in 
\eqr{e:sn yn}:
\begin{equation}\label{e:sg}
s_{j,n} = \rho_{j} - \rho_{0} \und  
y_{j,n} = (F(\rho_{j})  - \rho_{j}) - (F(\rho_{0}) - \rho_{0}), \quad (j=0,1,\dots,n-1).
\end{equation} 
The method we propose in the following section is centered on the 
current point
\begin{equation}\label{e:Pulay}
s_{j,n} = \rho_{j} - \rho_{n} \und  
y_{j,n} = (F(\rho_{j})  - \rho_{j}) - (F(\rho_{n}) - \rho_{n})\quad (j=0,1,\dots,n-1). 	
\end{equation} 

Many multisecant methods are easily understood by formulating the 
underlying optimization problem each of the approximate Jacobians 
(implicitly) solves.  We consider first the optimization problem
\begin{equation}\label{e:P1}
\ucmin{\frac{\alpha}{2}\|A-X\|^2 + \iota_C(X)}{X\in\Rbb^{k\times k}}
\end{equation}
where, throughout, the norm of a matrix is the {\em Frobenius} norm, 
$A$ is a real $k\times k$ matrix and $D,G$ are real $k\times m$ 
matrices such that the 
set $C$ defined below is nonempty: 
\begin{equation}\label{e:subspace}
C\equiv\left\{X\in \Rbb^{k\times k} \mbox{ such that }XD=G\right\}.
\end{equation}
 The corresponding indicator function of $C$, $\iota_C$, is defined by 
\[
\iota_C(X) = \begin{cases}0&\mbox{ if }X\in C \\
\infty & \mbox{ else.}
\end{cases}
\] 
The solution to the optimization problem \eqr{e:P1}  expressed as
\begin{equation}\label{e:prox1}
\argmin_X\left\{\frac{\alpha}{2}\|A-X\|^2 + \iota_C(X)\right\}
\end{equation}
is the {\em prox mapping}  \cite{Moreau62, Moreau65} 
of the  indicator function to $C$ at $A$:
\[
\prox_{1/\alpha,\iota_C}(A)\equiv 
\argmin_X\left\{\frac{\alpha}{2}\|A-X\|^2 + \iota_C(X)\right\}.
\]
It is an elementary exercise (see, for instance, \cite[Chapter 1, Section G]{VA}) 
to show that
\[
\prox_{1/\alpha,\iota_C}(A) = P_C(A)\qquad (\mbox{for all $\alpha>0$})
\]
where
\[
P_C(A) = \argmin_{X\in C}\{\|A-X\|\}
\]
is the projection of $A$ onto $C$.  
This projection has a simple closed form so long as the columns of $D$ are linearly independent:
\begin{equation}\label{e:P_C}
P_C(A) = A + (G-AD)\left(D^TD\right)^{-1}D^T.
\end{equation}

Specializing to multisecants, if $A=B_0$, the $n$'th 
approximation to the 
Jacobian, $D=S_n\in\Rbb^{k\times m}$ and $G=Y_n\in\Rbb^{k\times m}$ 
($1\leq m\leq n$), the columns of which are denoted  $y_j$ and $s_j$
respectively ($j\in[0,n]$), then we arrive at the multisecant 
extension of the 
Broyden's first update (MSGB) as studied by \cite{Barnes65, OrtegaRheinboldt70, 
GaySchnabel78, Schnabel83}:
\begin{equation}\label{e:MSGB}
B_{n+1} = P_C(B_0) = B_0 + 
(Y_n-B_0S_n)\left(S_n^TS_n \right)^{-1}S_n^T.
\end{equation}
Elementary calculations  using the Sherman-Morrison-Woodbury formula 
yield the multi-step recursion for $B_{n+1}^{-1}$ , analogous to \eqr{e:GBinv0}, 
\begin{equation}\label{e:MSGBinv0}
B_{n+1}^{-1} = B_{0}^{-1} + \left(S_n  - B_{0}^{-1}Y_n \right)
\Big( (S_n^TS_n)^{-1}S_n^TB_{0}^{-1}Y_n \Big)^{-1}(S_n^TS_n)^{-1}S_n^TB_{0}^{-1}
\end{equation}

For the update \eqr{e:Barnes} 
$v_n$ is the projection of $s_n$ onto the orthogonal complement of 
$[s_{n-m}, s_{n-m+1}, \dots, s_n]$ ($0\leq m< n$) where $s_j$ and $y_j$ are 
given by \eqr{e:sn yn}.  Here the update $B_{n+1}$ 
will satisfy $m+1$ {\em consecutive} secant equations 
$B_{n+1}s_j = y_j$ ($j=n-m,\dots, n$).  Methods based on this approach are 
called projected secant updates.  
This idea, however, seems not to have benefited from the endorsement of 
prominent researchers at the time, and hence there is little numerical experience to recommend it.  
A notable exception is the work of Martinez
and collaborators \cite{Martinez79, Martinez92, Martinez92b, Martinez00}.   
Sequences based on update \eqr{e:MSGB} are shown in \cite[Theorem 2.5]{Schnabel83} to be locally
q-superlinearly convergent if, in addition to other standard assumptions, the approximate Jacobians, 
$B_n$ stay close to the behavior of the true Jacobian, and if the columns of $S_n$ are strongly 
linearly independent as measured by the condition number 
\[
\kappa(S_n)\equiv \|S_n\|\left\|\left( S_n^TS_n \right)^{-1} S_n^{T}\right\|.
\]

An alternative specialization of \eqr{e:P_C} leads to a multisecant form of Broyden's second 
method (MSBB) if we let $A=H_0$, $D=Y_n$ and 
$G=S_n$ so that
\begin{equation}\label{e:MSBB}
H_{n+1} = P_C(H_0) = H_0 + 
(S_n-H_0Y_n)\left(Y_n^TY_n\right)^{-1}Y_n^T.
\end{equation}
To our knowledge, there are no published numerical comparisons of \eqr{e:MSBB} 
to alternatives, neither is there any published convergence 
theory, though we believe this
is only a minor modification of the theory for \eqr{e:MSGB}.  

Yet another specialization of \eqr{e:P_C} is a 
simplex gradient technique \cite{Mifflin75} 
for vector-valued functions proposed by 
Gragg and Stewart \cite{GraggStewart76} and 
implemented in \cite{Martinez79}.  Here the Jacobian update is given by
\begin{equation}\label{e:simplex gradients}
B_{n+1} = P_C(0) = Y_n(S_n^TS_n)^{-1}S_n^T.
\end{equation}
where $A=0$, 
$D=Y_n$ and $G=S_n$ where the steps $s_n$ and residual differences $y_n$
are centered on the {\em initial point} $\rho_0$ as in \eqr{e:sg} . 

Independent studies appearing in the physics literature that parallel the 
mathematical  literature have a different
variational form.  The various approaches can all be shown to be
specializations of the the following optimization problem
\begin{equation}\label{e:P2}
\ucmin{\frac{1}{2}\sum_{j=1}^n\alpha_j\dist_{C_j}^2(X)+
\frac{\alpha_0}{2}\|A-X\|^2}{X\in\Rbb^{k\times k}}
\end{equation}
where  each $C_j$ is a set of the form \eqr{e:subspace} and $A\in\Rbb^{k\times k}$, 
and $\dist_{C_j}(X)$ is the Euclidean distance
of $X$ to the set $C_j$.  
A short calculation  yields  the solution $X_*$ to \eqr{e:P2}
\begin{equation}\label{e:prox dist}
X_* = \frac{\alpha_0}{\sum_{j=0}^n\alpha_j}A +
\sum_{j=1}^n\frac{\alpha_j}{\sum_{j=0}^n\alpha_j}P_{C_j} A.
\end{equation}
From \eqr{e:P_C} this can be written explicitly as 
\begin{equation}\label{e:prox dist2}
X_* = 
\sum_{j\in\Jbb}\gamma_j A
+\sum_{j=1}^n\gamma_j
\left( (G_j-AD_j)\left(D_j^TD_j\right)^{-1}D_j^T\right)
\end{equation}
where 
\begin{equation}\label{e:gamma_j}
\gamma_j\equiv\frac{\alpha_j}{\sum_{j=0}^n\alpha_j}.
\end{equation}

Specializing to multisecants, let $A=B_n$, $D_j=s_j$ and 
$G=y_j$, where
$s_j$ and $y_j$ are defined by \eqr{e:sn yn}.   Then the optimization problem \eqr{e:P2}
corresponds to the variational formulation of a method proposed by  
Vanderbilt and Louie \cite{VanderbiltLouie84}.  
A local convergence analysis, together
with numerical tests are studied in \cite{CrittinBierlaire03}.  
Our derivation and formulation of the update, however, 
appears to be new and clarifies the connections between their method and 
\eqr{e:MSGB} above:
\begin{equation}
\label{e:Vanderbilt}
B_{n+1} = \sum_{j=0}^n\gamma_j B_n
+\sum_{j=1}^n\gamma_j
\left( (y_j-B_ns_j)\left(s_j^Ts_j\right)^{-1}s_j^T\right).
\end{equation}
If instead we let let $A=H_n$, $D_j=y_j$ and 
$G_j=s_j$, we get the update proposed by Johnson 
\cite{Johnson88}:
\begin{equation}
\label{e:Johnson}
H_{n+1} =
\sum_{j=0}^n\gamma_j H_n
+\sum_{j=1}^n\gamma_j
\left( (s_j-H_ny_j)\left(y_j^Ty_j\right)^{-1}y_j^T\right).
\end{equation}
Again, our derivation is different, and the new formulation makes the 
connection with \eqr{e:MSBB} more transparent.  

The weighting scheme of \cite{VanderbiltLouie84,Johnson88}
is similar to a technique proposed by Pulay \cite{Pulay80}.  
A dynamic weighting scheme that optimizes the weights $\gamma_j$ simultaneously
with the determination of the matrix $H_n$ or $B_n$ is possible via the 
extended least squares techniques outlined in \cite{Luke02a}.  
From the analysis above, the ``active ingredients'' in the methods of 
\cite{VanderbiltLouie84} and \cite{Johnson88} involve the 
nonuniform weighting of the 
conventional multisecant formulations of Broyden's 
methods, and the centering of the prox mapping on the most recent
matrix approximation $B_n$ and $H_n$ rather than on 
$B_0$ and $H_0$ as is the convention in \eqr{e:MSGB} and \eqr{e:MSBB}.
A variation of \eqr{e:Johnson} due to 
Kawata et al \cite{Kawata98} combines 
the method of Johnson with a construction of the columns of $S_n$ and
$Y_n$ proposed by Pulay \cite{Pulay80} and given in \eqr{e:Pulay}. 
We note that the methods summarized by \eqr{e:Vanderbilt}-(\ref{e:Johnson}) and their
relatives solve single matrix secant equations {\em in parallel} while the methods summarized by 
\eqr{e:MSGB}-(\ref{e:MSBB}) solve multiple matrix secant equations {\em simultaneously}, 
which is more restrictive.  

In the above analysis we are not specific about {\em how many} 
previous steps should be included in the matrices $S_n$ and $Y_n$.  Recall that these matrices
are made up of $m$ columns of previous step information where $m\in[1,n]$.  If $m<n$ then
one is implicitly executing a {\em limited memory} technique.  The name limited memory 
stems from the sequential ordering of steps according to \eqr{e:sn yn} and refers
to the practice of excluding points more than $m$ steps previous in the construction of  
the current matrix update \cite{ByrdNocedalSchnabel94}.
If one constructs $S_n$ and $Y_n$ via \eqr{e:Pulay}, as we do in the following numerical experiments,
then one would exclude points that are most distant from the current point $\rho_n$.  
This is a reasonable strategy for highly nonlinear problems, where the linear approximation that
is at the heart of quasi-Newton methods is only valid on a local neighborhood of the 
current point.  For extremely large problems such a strategy is also expedient since the matrix updates 
need not be explicitly stored as they can be constructed from a few stored vectors.  

Finally, we note some other multisecant approaches that
don't fit into the framework above.  In \cite{Schnabel83} a multisecant BFGS 
update is studied, however there do not appear to be computationally 
efficient and stable methods for 
generating updates that preserve the algebraic structure of the original BFGS 
update, namely symmetry and positivity.  The SR1 update,  also studied in 
\cite{Schnabel83}, is shown to be computationally efficient and stable though 
at the cost of symmetry. In contrast to these, Broyden's updates are natural 
candidates for multisecant approaches since they do not enforce symmetry 
of the Jacobian  (they are designed for solving systems of nonlinear equations 
rather than for solving optimization problems as with BFGS and SR1).  
For problems where the Jacobian is ill-conditioned or singular at the solution,
Frank and Schnabel \cite{Schnabel84, Schnabel86} consider methods
based on the third terms of the Taylor series expansion in \eqr{e:Taylor}, 
which they called ``tensor methods'', that build upon the multisecant model. 
 They reported on average improvements of $33\%$ for highly controlled 
test problems in which the null space of the Jacobian has dimension $1$ or 
$2$.  There appears to be little more numerical experience with tensor methods 
for matrix secant approaches.  While this may be an avenue for further 
research, we do not pursue this idea in the present work.  

\section{\label{s:new stuff} New Developments: safeguarded multi-secants}
Newton-like algorithms, such as Broyden's methods, are not global techniques for solving 
equations and can behave wildly, even chaotically, far from a solution.  As we have 
seen in the previous section, solving the self-consistent field equations is equivalent 
to finding the fixed points of the SCF mapping  $F$ given by \eqr{e:fix}.  
The extent to which many algorithms behave, or misbehave, depends on the functional 
properties of the SCF mapping.  Consider, for instance, an even simpler algorithm to 
Broyden's method, namely
\begin{equation}\label{e:banach}
\rho_{n+1} = F(\rho_n);
\end{equation}
that is, we replace the current density with the density under the SCF mapping.  
This is a simple form of a fixed point iteration.  If $F$ is a {\em contraction} on 
some closed subset of the space of densities (i.e. points move {\em closer} to one another 
under the mapping $F$), then the sequence $\rho_{n}$ converges to the {\em unique} 
fixed point $\rho_*$ of $F$ (this is known as the Banach Contraction Theorem – see, 
for instance \cite{BorLew06}).  If $F$ is not a contraction,  then \eqr{e:banach} 
could continue forever without ever approaching a fixed point. Successive iterates might 
form a characteristic path, or they might behave chaotically.  
One could reasonably conjecture that many Kohn-Sham SCF mappings are not everywhere 
contractive since \eqr{e:banach} is not a popular numerical method.  Less restrictive than 
contractive mappings are {\em nonexpansive} mappings (i.e. points do not 
move further apart under the mapping $F$).   Of course, all contractions are nonexpansive, 
but the converse does not hold.  If $F$ is a nonexpansive mapping on a closed, 
bounded and {\em convex} subset of the space of densities, then the modified mapping 
$\widetilde{F}(\rho)\equiv \rho+ \lambda(F(\rho)-\rho) $ for small 
$\lambda>0$ is contractive, and $\widetilde{F}(\rho)$, and hence $F$, has a fixed point 
\cite{Browder65,Kirk65}.  The iteration 
to find the fixed point of $\widetilde{F}$ corresponding to \eqr{e:banach} in terms of the 
original mapping $F$ is 
\begin{equation}\label{e:Pratt}
\rho_{n+1} = \widetilde{F}(\rho_n)\equiv \rho_n+ \lambda(F(\rho_n)-\rho_n).
\end{equation}
Most readers will recognize this as the Pratt step \cite{Pratt}.  
Convergence is guaranteed for nonexpansive
SCF mappings on compact convex regions (though convergence can be extremely slow), but 
if $\lambda$ is too large, or 
$F$ is only locally nonexpansive, or worse, not even nonexpansive, then the
theory for fixed point iterations like \eqr{e:Pratt} and even Broyden's methods does not have 
much to say and numerical behavior cannot be guaranteed.  In other words, anything can happen.  

In particular, since it is possible for the SCF mapping to have several fixed points there is no 
guarantee that an algorithm will converge to the {\em correct} fixed point, as is the case if 
the fixed point is nonphysical, known as a ``ghost band''.  
Indeed, the iteration need not converge at all.  Even if one supposes that the SCF mapping
is contractive on closed neighborhoods of each of these fixed points, if an algorithm takes too large
a step it could leave the domain of attraction of one fixed point and drift towards another fixed
point.  This process of drifting between attractors of the SCF mapping could continue 
indefinitely, and with a  chaotic trajectory, if the algorithm is not sufficiently well controlled.

Note that one can extend the above concepts to a subset of the density
variables.  For instance, the $sp$-electron states might behave well, while
$d$-electron states might be very difficult to converge.
This suggests a certain separability of the variables associated with
these quantities, and that the SCF mapping is more sensitive to some subsets of the density
variables than others.  Indeed, a frequent observation exhibited in Section \ref{s:results} is
that the density within the muffin-tins often behaves very differently to the density in the
interstitial region.

The problematic part of the Kohn-Sham mapping is the effective potential
$V_\rho$.  In general, there is no closed form for $V_\rho$.  For certain approximations, denoted 
$\widetilde{V}$,
it is possible to prove the correspondence between the fixed points of the corresponding SCF mapping 
$F_{\widetilde{V}}$ and solutions to the Kohn-Sham equations \cite{Prodan05}, and, moreover, that 
$F_{\widetilde{V}}$ is a contraction \cite{Prodan03}.  However, for exact $V_\rho$ at finite 
temperatures existence and uniqueness of fixed points is an open question, further complicated 
by the occurrence of systems with multiple coexisting phases \cite{Prodan05}.   For the 
practitioner who simply wants her software to converge for a particular example, unfortunately
this means that the algorithms come only with extremely limited warranties that may not even 
be verifiable.  

With this caveat in mind, and before we present the details of our numerical method, we take 
a moment to describe in physical terms some of the features of ab-initio calculations that are 
problematic, together with common symptoms of poorly convergent problems. 
\begin{enumerate}
\item\label{i:easy} In many cases, for instance bulk MgO, the algorithms reach an acceptable solution 
in a surprisingly small number of iterations, e.g. $10-20$ for $10^4$ unknown density 
components.   This implies that, at least for a substantial subset of the density parameters, the 
domain of attraction of the fixed point is large and the SCF mapping has ``good'' functional
properties  on this domain.

\item\label{i:scaling} 
In some cases there can be issues with the scaling of different parts of the density 
because they are represented in quite different fashions. For instance, with LAPW methods 
the plane wave components outside the muffin-tins are represented by the Fourier 
coefficients whereas the density inside the muffin-tins is expanded in terms of spherical
harmonics.

\item\label{i:Muffins}
The conventional wisdom for LAPW-based methods is that the muffin-tins should be as large 
as possible without overlapping.  This implies that the basis set used for the muffin-tins is 
somehow better suited for the physics or for the geometry of the atoms.  This is manifested in 
more rapid convergence of the coefficients corresponding to these basis elements and indicates
that the domain of attraction of the fixed point for these coefficients is large relative to the 
domain of attraction for the fixed point of the plane wave elements.

\item\label{i:hard}  
The most physically interesting problems are often harder to solve.
A spin unpolarized DFT calculation of NiO, for example, may converge very slowly. The slow 
convergence of the mixing cycle is in part because spin unpolarized the system is metallic, 
but is also coincidental with an imperfect functional description of this system, in which case 
the Hamiltonian in \eqr{e:fix} can be ill-posed.  It is not 
uncommon to compromise on the physical model, particularly for large and complicated 
problems. 

\item\label{i:sloshing}  In some cases, for instance when there are $d$ or $f$ electrons,
charge carriers are in a large unit cell and for surfaces, mixing converges poorly and can 
easily diverge.  In the literature this is called ``charge 
sloshing'' because one has oscillations of charge density between different spatial regions of the 
problem or between different local states such as $d$-electrons. Mathematically this sometimes corresponds 
to ill-conditioning when a small change in the density $\rho$ can lead to large change in $F(\rho)$,
with large eigenvalues of the matrix $H$ (or small eigenvalues of $B$). Alternatively, it may be that
the higher-order terms of \eqr{e:Taylor} are large, so neglecting them is only appropriate for a
very small change in the density.  A third possibility in the case of charge sloshing is that the 
SCF mapping is not nonexpansive (and hence not contractive) along this trajectory.

\item\label{i:ghost band} 
It is possible with a LAPW method for the calculations to enter regions where the new density has become 
unphysical.  Such numerical densities are called ``ghost bands'' and are unphysical solutions of the 
Kohn-Sham equations \eqr{e:fix}. 
\end{enumerate}

\begin{figure}
\begin{center}
(a)\includegraphics[height=7cm,width=7cm,angle=0]%
         {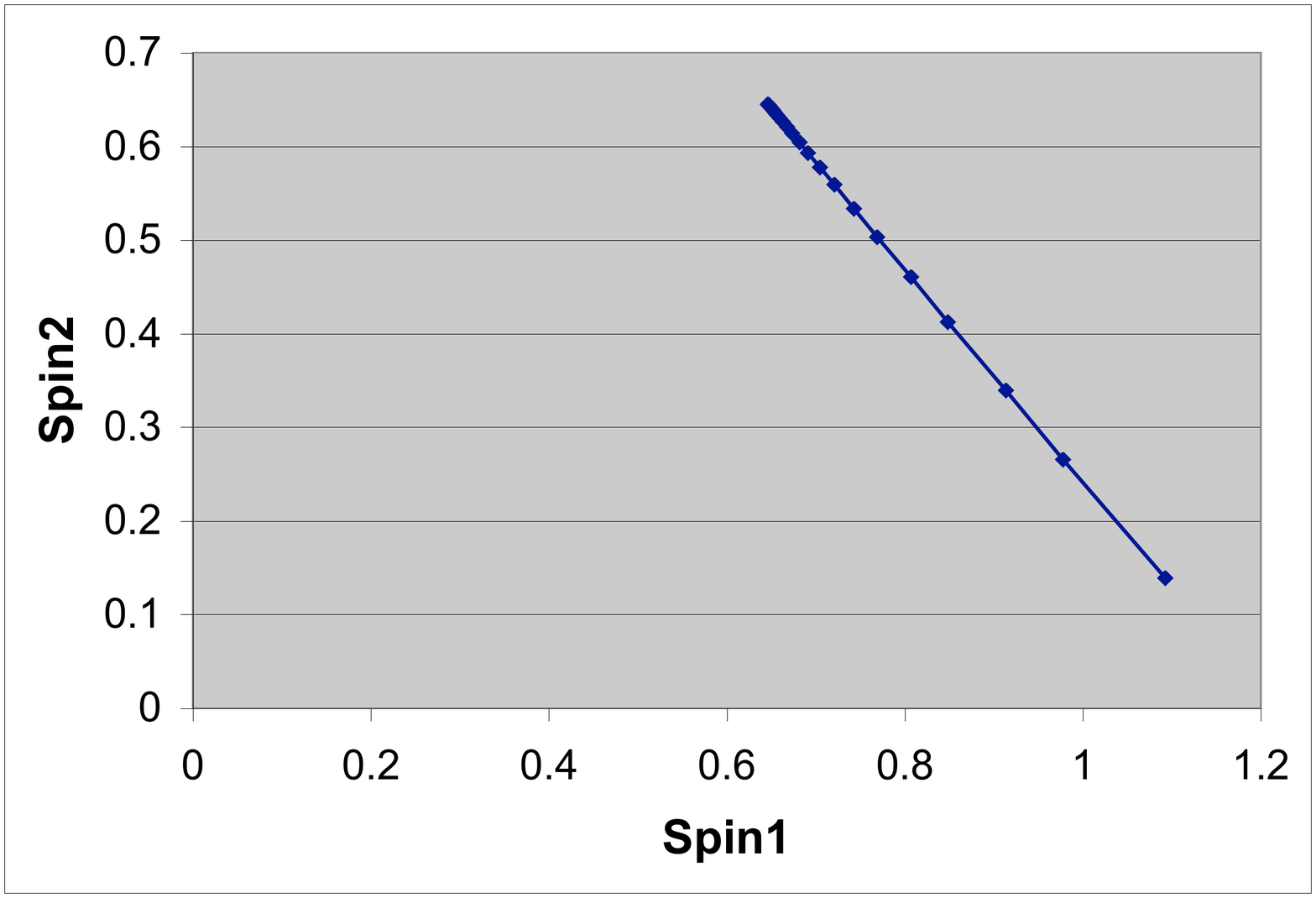}
\hfill
(b)\includegraphics[height=7cm,width=7cm,angle=0]%
         {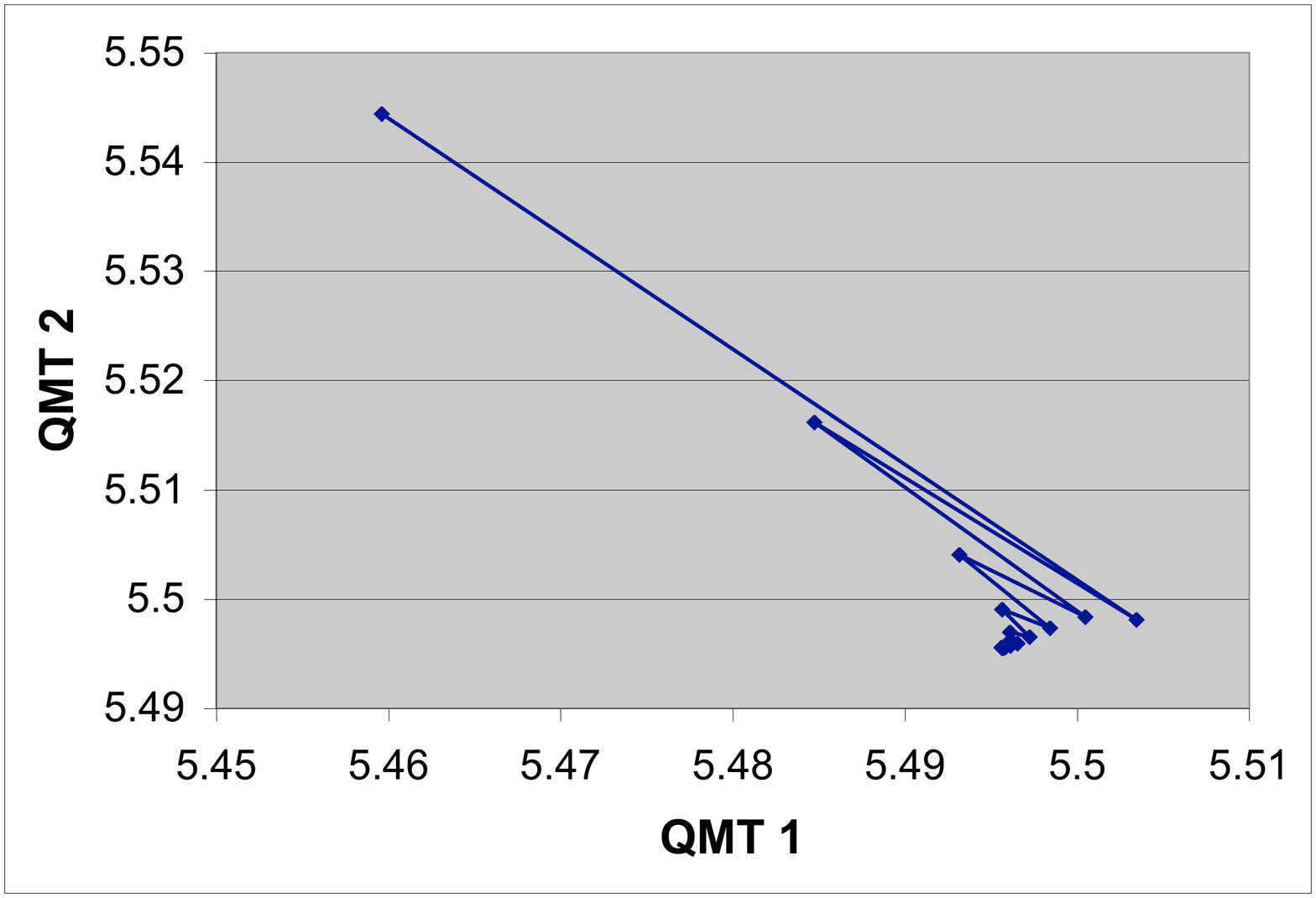}\\
\caption{\label{f:Dali}  Iterates for an $O_2$ molecule
with atomic densities having a spin of $+2$ the other with $0$.
Figure (a) shows the spin within the muffin tins and (b) shows the
total charge within the muffin tins (QMT)  during a
representative SCF iteration.  
}
\end{center}
\end{figure}

To illustrate these features a simple model is an $O_2$ molecule
starting from atomic densities where the two atoms are deliberately
treated differently, one starting with a spin of $+2$ the other with $0$.
Shown in \figr{f:Dali} 
is the variation of the spin (a) and
total charge within the muffin tins (QMT) (b) varies during a
representative SCF iteration. While the spins converge 
smoothly to the final solution, the total charge oscillates or ``sloshes''.
\figr{f:Dali} shows the behavior of the spin (a) and the 
total charge, QMT (b), within the muffin tins for sequences of densities 
approaching the solution.
For this sequence of points the spin of the density $\rho_n$, shown in \figr{f:Dali}(a), 
converges smoothly to the solution.  A view of the charges of the electrons within the 
muffin tins, \figr{f:Dali}(b), tells a different story.  The total charge
within the muffin tins appears to zig-zag toward the solution.

The behavior of different physical quantities for sequences of points approaching 
a fixed point is independent of algorithm design
(as opposed to algorithm performance) and is symptomatic of the functional properties of 
the SCF mapping, i.e. the underlying physical problem.  Our purpose is to design an 
algorithm that performs well for both 
regular SCF mappings as well as more pathological cases.  
As each model presents unique analytical irregularities,  we seek to address
the most common challenges in our numerical algorithms.  

%

\subsection{\label{s:heuristics} Toward a robust algorithm}
In addition to numerical performance for a wide range of DFT calculations, a good numerical 
method, in our opinion, will require little or no user intervention.  A key observation is that
the dimension of the underlying problem is on the order of $10^4$ 
or higher while the information that one uses to model the fixed point mapping is at most dimension 
$2n$ where $n$ is the number of iterations (on the order of $10^0$).  
As mentioned above, the conventional view is that the $n$ steps and residual differences generated 
in matrix secant methods are deterministic points on a path to the solution.  
While this is true for problems of small and moderate dimension, 
for high-dimensional problems such as SCF calculations, we  can alternatively 
consider the $n$ steps as random samples of a high-dimensional mapping,
albeit with decreasing randomness as the matrix secant model is refined in subsequent iterations.
We therefore consider the vectors $s_j$ and $y_j$ given by \eqr{e:sn yn} as merely data samples with
{\em less} significance given to the order in which the samples were collected than with conventional matrix 
secant approaches.  In particular, we center our model on the current iterate and calculate
all steps and residual differences computed relative to the current point according to 
\eqr{e:Pulay}. 
The algorithm then predicts the behavior of the SCF mapping \eqr{e:fix} 
at $\rho_n$  given an appropriate model of the previous 
data. 

The multi-secant methods detailed in the previous sections can all be rewritten as
\begin{eqnarray}\label{e:gen_heur}
\rho_{n+1} &=& \rho_n - H_0\Big(g_n-Y_{n-1}A_ng_n\Big) - S_{n-1}A_ng_n.
\end{eqnarray}
where $g_n = F(\rho_n)-\rho_n$, $A_n$ is a matrix dependent on the method, and $H_0$ 
is an initial matrix secant estimate.  Let us  write this as
\begin{equation}\label{e:Fn}
	s_n = u_{n}+p_{n}
\end{equation}
where, according to \eqr{e:gen_heur}, $p_{n}=-S_{n-1}A_ng_n$ and 
$u_{n}=-H_0(I-Y_{n-1}A_n)g_n$.  We show in Subsection \ref{s:details} that
$p_n$ is the part of the vector $g_n$ that can be explained by (is in the range of) 
the data at step $n$, and $u_n$ is the part that is orthogonal to this information, 
and hence unpredicted.  
The numerical challenges described in (\ref{i:hard})-(\ref{i:scaling}) are embedded in the 
unpredicted component of the new step.  Written this way, we can distinguish two numerical 
challenges. First and most obvious, one must safeguard against the unpredicted behavior $u_n$.  
More subtle is the \textit{a posteriori} accounting of the
vector $g_n$ and an underlying {\em distance to ill-posedness} of the 
prior information, which reflects the quality of the 
information.  In other words, there is {\em no free lunch}:  not only do we have
to be cautious of unpredicted behavior, but we also have to be careful of how we treat our
predicted behavior.  It is reasonable to expect that with more information the size of the 
unknown orthogonal component $u_n$ will decrease, however, this is not always the 
case if the model does not adequately capture the true SCF mapping, or 
does not adequately adjust for the presence of multiple scales within the model.  
In addition, a poorly conditioned, or nearly ill-posed model might have an 
unstable direction in which numerical errors overwhelm meaningful information 
on the same scale.  
Our numerical strategy must safeguard against both large 
unpredicted components and unstable models.  We show next
how this perspective translates into algorithms.

\subsection{\label{s:details} Scaling, regularization, and preconditioning}
%
The discussion in the previous subsection is easily made rigorous when we consider the 
multisecant formulation of Broyden's second method \eqr{e:MSBB} MSBB with the steps 
centered according to 
either \eqr{e:sn yn} or \eqr{e:Pulay}.  To see this, note that from 
\eqr{e:rho_n+1 backward} with $H_{n}$ replaced by \eqr{e:MSBB}, 
the modification of \eqr{e:gen_heur} to MSBB amounts to letting 
\begin{equation}\label{e:heur1}
 A_n\equiv\left(Y_{n-1}^TY_{n-1}\right)^{-1}Y_{n-1}^T.
\end{equation}  
Note that
$\left(Y_{n-1}^TY_{n-1}\right)^{-1}Y_{n-1}^Tg_n$ is the solution to the least squares 
minimization problem
\begin{equation}\label{e:lsY}
\ucmin{\tfrac12 \|Y_{n-1}z - g_n  \|^2}{z\in\Rbb^{m}},
\end{equation}
where $m\in[1,n-1]$ is the number of previous data points used in the update.  
In other words, $\left(Y_{n-1}^TY_{n-1}\right)^{-1}Y_{n-1}^Tg_n$ is the element 
in the domain of $Y_{n-1}$ that comes closest 
(in the least squares sense) to ``predicting'' the vector $g_n$.  
It follows, then, that 
\begin{equation}\label{e:orthog1}
\Big(I-Y_{n-1}A_n\Big)g_n = \Big(I-Y_{n-1}\left(Y_{n-1}^TY_{n-1}\right)^{-1}Y_{n-1}^T\Big)g_n
\end{equation}
is the orthogonal projection 
of $g_n$ onto the space orthogonal to the residual
differences $y_j$ defined by one of \eqr{e:sn yn} or \eqr{e:Pulay}, our prior data.   


The above discussion assumes that $Y_{n-1}$ is full-rank.  If the columns
of $Y_{n-1}$ are nearly linearly dependent, then the inverse $\left(Y_{n-1}^TY_{n-1}\right)^{-1}$
can be numerically unstable.  More fundamentally, we are implicitly assuming that the 
approximation to the Jacobian in \eqr{e:rho_n+1} is, first of all, valid on the neighborhood of 
$\rho_n$ defined by the other data points and, second of all, that the Newton step is the 
{\em right} step to take.  If either one of these assumptions does not hold, as would be the 
case when we are far from the solution and our sample points are far apart, conventional 
optimization strategies link local and global techniques by 
allowing steps to rotate between the steepest descent direction (in the present setting, 
the direction of the vector $g_n$) and a Newton-like direction.
One well-known strategy of this kind is the Levenberg-Marquardt algorithm
\cite{Levenberg, Marquardt}.  We propose
a different technique that is an unusual use of a classical regularization technique usually 
attributed to Tikhonov  \cite{Tikhonov1,Tikhonov2,Hansen98} and rediscovered in the statistics 
community under the name of ridge regression \cite{Hoerl}, though the more general notion of 
proximal mappings due to Moreau \cite{Moreau62} predates both of these.  
In particular we regularize \eqr{e:lsY} in the usual way:
\begin{equation}\label{e:lsY_reg}
\ucmin{\tfrac12 \|Y_{n-1}z - g_n  \|^2 +\tfrac{\alpha}{2}\|z\|^2}{z\in\Rbb^{m}},
\qquad(\alpha>0).
\end{equation}
Recalling the prox mapping introduced in the previous section, we could center
the regularization more generally on any point $z_0$, however, since we have
no prior information about $z$ the logical choice for the regularization is
to center it on the origin.  The solution to \eqr{e:lsY_reg} is 
\begin{equation}\label{e:Tikhprox}
z_n = \left(Y_{n-1}^TY_{n-1}+\alpha I\right)^{-1}Y_{n-1}^Tg_n
\end{equation}
which yields the following regularization of $A_n$:
\begin{equation}\label{e:heur1_reg}
A_n^\alpha\equiv \left(Y_{n-1}^TY_{n-1}+\alpha I\right)^{-1}Y_{n-1}^T.
\end{equation}
Note that as $\alpha\to \infty$, $A_n^\alpha\to 0$, and the step 
generated by \eqr{e:gen_heur} rotates to the direction $H_0g_n$.
If $H_0$ is a scaling of the identity (see Subsection \ref{s:step}) then in the context
of optimization, where $g_n$ is actually a gradient, the stronger 
the regularization, the more the step rotates in the direction of steepest 
descent and away
from a Newton-like direction.  We thus interpret the regularization 
parameter in both the conventional way, stabilizing 
$\left(Y_{n-1}^TY_{n-1}\right)^{-1}$, and 
as an estimation of the uncertainty of the approximate Newton step.   
Given our understanding
of the previous step data as pseudo-random samples from an unknown process, 
the latter interpretation has a very natural explanation in terms of the Wiener
filter for a signal with normally distributed zero-mean white noise.  The
size of the regularization parameter corresponds to the energy of the noise, or uncertainty
in our model. 

Johnson \cite{Johnson88} proposes a scaling of the columns of the matrices
of $Y_n$ and $S_n$ for numerical reasons, though this can easily be shown
to have no formal impact on the algorithm.  In the context of regularization, 
however, such a scaling  can have a significant effect on the 
choice of the regularization parameter.  This scaling  is equivalent 
to multiplication of the matrices $Y_n$ and $S_n$ on the right by the 
diagonal matrix $\Psi_n$.   The rescaled least squares problem analogous to 
\eqr{e:lsY_reg} is 
\begin{equation}\label{e:lsYPsi_reg}
\ucmin{\tfrac12 \|Y_{n-1}\Psi_nz - g_n  \|^2 +\tfrac{\alpha}{2}\|z\|^2}{z\in\Rbb^{m}},
\qquad(\alpha>0)
\end{equation}
with the solution 
\[
\left(\Psi_nY_{n-1}^TY_{n-1}\Psi_n+\alpha I\right)^{-1}\Psi_nY_{n-1}^T.
\]
It follows immediately from this that if we normalize the columns of 
$Y_{n-1}$, 
\begin{equation}\label{e:Psin}
\Psi_n =\left(\begin{array}{cccc}1/\|y^{(n-1)}_1\|&0&\dots&0\\
0&1/\|y^{(n-1)}_2\|&\ddots&\vdots\\
\vdots&\ddots&\ddots&\vdots\\
0&\dots&0&1/\|y^{(n-1)}_m\|
\end{array}\right)
\end{equation}
where $y^{(n-1)}_j$ is the $j$-th column of $Y_{n-1}$,
then our regularization parameter will be {\em independent} of scaling between 
the columns of the matrix $Y_{n-1}$.   This allows for more robust regularization strategies.  
Viewing the regularization as a Wiener filter applied to the approximate
Newton step, the scaling reduces the effect of outliers on the regularization parameter 
in the least squares estimation, these outliers coming from steps that are relatively far 
away from the solution we seek.  
We denote the matrix corresponding to this scaling, together with the regularization
$\alpha$ by $A_n^{\alpha,\Psi}$ where
\begin{equation}
A_n^{\alpha,\Psi}\equiv \Psi_n\left(\Psi_nY_{n-1}^TY_{n-1}\Psi_n+\alpha I\right)^{-1}\Psi_nY_{n-1}^T.
\label{e:heur1_reg-scale}
\end{equation}
The step is then generated by \eqr{e:gen_heur} with $A_n$ replaced by $A_n^{\alpha,\Psi}$.

Formalizing our approach for Broyden's first method MSGB is not as obvious.  
From 
\eqr{e:rho_n+1 forward} with $B_{n}^{-1}$ replaced by \eqr{e:MSGBinv0}, 
the modification of \eqr{e:gen_heur} for MSGB amounts to letting $H_0=B_0^{-1}$ and 
\begin{equation}\label{e:heur2}
 A_n\equiv \Big(\left(S_{n-1}^TS_{n-1}\right)^{-1}
S_{n-1}^TB_0^{-1}Y_{n-1}\Big)^{-1}
\left(S_{n-1}^TS_{n-1}\right)^{-1}S_{n-1}^TB_0^{-1}.
\end{equation}  
Again, we note that $\left(S_{n-1}^TS_{n-1}\right)^{-1}S_{n-1}^TB_0^{-1}w$ is the solution to 
the least squares problem
\[
\ucmin{\tfrac12 \|S_{n-1}z - B_0^{-1}w  \|^2}{z\in\Rbb^{n-1}}.
\]
If, in addition, $B_0=1/\sigma I$, then an elementary calculation yields
the simplification to \eqr{e:heur2}
\begin{equation}
 A_n= \left(S_{n-1}^TY_{n-1}\right)^{-1}S_{n-1}^T
\label{e:heur2b}
\end{equation}
If $\left(S_{n-1}^TY_{n-1}\right)^{-1}$ is well-defined, then the mapping 
$I-Y_{n-1}A_n$ is a 
{\em nonorthogonal
projection}\footnote{A projection is defined as any mapping $P$ such that $P^2=P$.  An
orthogonal projection onto a set $C$ is the point in $C$ that is nearest, with respect to the norm, 
to the point being projected.}
onto the nullspace of the columns of $S_{n-1}$, or in other words, a projection onto 
the space orthogonal to the range of the columns of $S_{n-1}$, our prior step data.
Unlike \eqr{e:heur1} the projection is {\em not} to a nearest element in the range of
 $S_{n-1}^\perp$, hence, by definition, the resulting step will be larger than the orthogonal 
projection.  

The above assumes that $\left(S_{n-1}^TY_{n-1}\right)^{-1}$ is well-defined, which
is likely, but this says nothing of whether or not $S_{n-1}^TY_{n-1}$ is 
well-conditioned.   Indeed, $S_{n-1}^TY_{n-1}$ need not have real eigenvalues, or
be positive definite.  Regularization of 
$\left(S_{n-1}^TY_{n-1}\right)^{-1}$ in \eqr{e:heur2}  gives 
$\left(S_{n-1}^TY_{n-1}+ \alpha I\right)^{-1}$
which shifts the eigenvalues to the right.  Since $S_{n-1}^TY_{n-1}$ could have 
negative eigenvalues, unless $\alpha$ is chosen large enough, 
this regularization could result in an even {\em more} ill-conditioned matrix.  
Our numerical experience is that  $\alpha>10^{-6}$ is sufficiently large to avoid 
this possibility for the applications of interest to us.  

We turn next to preconditioning and scaling. We propose rescaling the density $\rho_n$ to 
account for multiple scales between the interstitial electrons and the muffin tin 
electrons.  
There are many possible strategies.  Such scalings are generically 
represented by multiplying the density 
$\rho_n$ at each iteration $n$
{\em on the left} by an arbitrary invertible diagonal matrix $\Omega_n$.   The same scaling 
must also be applied to the result of the SCF mapping acting upon $\rho_n$.  
If these scalings 
are applied {\em before} the mixing operation and undone after the mixer yields
its proposed step, then one need not change any of the formalism above;  specifically, 
one replaces $Y_n$, $S_n$, and $A_n$ in \eqr{e:gen_heur} with  
$\Yhat_n\equiv\Omega_n Y_n$,  $\Shat_n\equiv\Omega_n S_n$, and,
\begin{equation}
A_n^{\alpha,\Psi_n,\Omega_n}\equiv \begin{cases}
\Psi_n\left(\Psi_n\Shat_{n-1}^T\Yhat_{n-1}\Psi_n+\alpha I\right)^{-1}\Psi_n
\Shat_{n-1}^T\Omega_n&\mbox{(MSGB), or}\\
\Psi_n\left(\Psi_n\Yhat_{n-1}^T
\Yhat_{n-1}\Psi_n+\alpha I\right)^{-1}
\Psi_n\Yhat_{n-1}^T\Omega_n 
&\mbox{(MSBB)}
\label{e:heur1_regb}
\end{cases}
\end{equation}
%
One of the benefits of such scalings is to increase the numerical 
accuracy of matrix multiplication when the matrices consist of elements of 
vastly different scales.

The preconditioner used in the numerical experiments in Section \ref{s:results} 
rescales the change in the interstitial electrons relative to that in the 
muffin-tin electrons.   Recall that the residual of the SCF mapping 
for the density $\rho_n$ is $g_n=F(\rho_n)-\rho_n$.  We represent the interstitial and 
muffin-tin portions of this residual by $g^{(I)}_{n}$ and  $g^{(M)}_{n}$ respectively where
\[
g_n=\left(\begin{array}{c} g^{(I)}_{n}\\ g^{(M)}_{n}\end{array}\right).
\] 
Denote the averages of the residuals of these components separately by 
\begin{equation}\label{e:g^ave}
\gbar^{(I)}_n =  \sum_{j=0}^n\|g^{(I)}_{j}\|/\|g_{j}\|,
\und 
\gbar^{(M)}_n =  \sum_{j=0}^n\|g^{(M)}_{j}\|/\|g_{j}\|, 
\end{equation}
Using the 
formalism above, our preconditioner $\Omega_n$ is defined by 
\begin{equation}\label{e:precond}
\Omega_n=\left(\begin{array}{cc}\omega_n I_1& 0\\ 0& I_2   \end{array}\right)
\where
\omega_n=\sqrt{\frac{\gbar^{(M)}_n}{\gbar^{(I)}_n}}
\end{equation}
 and $I_j$ is the $l_j\times l_j$ identity matrix where $l_j$ is the dimension of the 
interstitial/muffin tin electrons respectively. 
We note that the $\omega_n$ term enters the multisecant form squared,
hence our use of a square root.  Removing this square root is also reasonable, and in some 
cases is better in numerical tests, but it can be less stable and lead to runaway behavior 
where the interstitial regions converge too rapidly, upsetting the balance between these and 
the muffin tin electrons.   More sophisticated preconditioning are also plausible, for 
instance a dielectric term for the plane waves 
\cite{Dielectric1, Dielectric2}, though we found this simple form to be very effective. 

Before concluding this section we address a physical point.  A relevant question 
is whether the step constructed by these matrix secant methods 
conserves the total charge --  if not, then an additional 
constraint is needed.  By construction all the $s_n$, $y_n$ values conserve charge,
 as does $g_n$, and the
preconditioners and scalings simply change the matrix given by \eqr{e:heur1_regb}. 
The result will then conserve charge automatically within numerical accuracy, so no 
explicit charge constraint is necessary.

\subsection{\label{s:step} Step Control}

In Broyden's original numerical experiments he constructed 
$B_{0}$ from a finite difference approximation to the true Jacobian 
(see \cite[Section 7]{Broyden65}).  This is not a practical approach for extremely large problems such
as DFT calculations.  For large problems, the convention for the initial estimate 
$B_{0}$ or $H_{0}$ is a scaling 
of the identity; that is, at each iteration $n$ we choose 
$H_{0,n}=\sigma_n I$ and likewise $B_{0,n}=1/\sigma_n$. 
The choice of the scaling is critical -- if it is poorly chosen iterations can stagnate 
or diverge.  Note that the generating matrix can vary at each iteration.  
With this generating matrix the unpredicted component of the step 
$s_n$, given by \eqr{e:gen_heur}-(\ref{e:Fn}) is 
$u_{n}=-\sigma_n(I-Y_{n-1}A_n)g_n$.  The scaling $\sigma_n$ has no impact 
on the predicted component $p_n=-S_{n-1}A_ng_n$.  A technical discussion of 
strategies for choosing $\sigma_n$ are intimately connected to a convergence 
analysis of the algorithms, which is the topic of subsequent work.  

\begin{figure}\begin{center}
(a) \includegraphics[height=5cm,width=6cm,angle=0]%
         {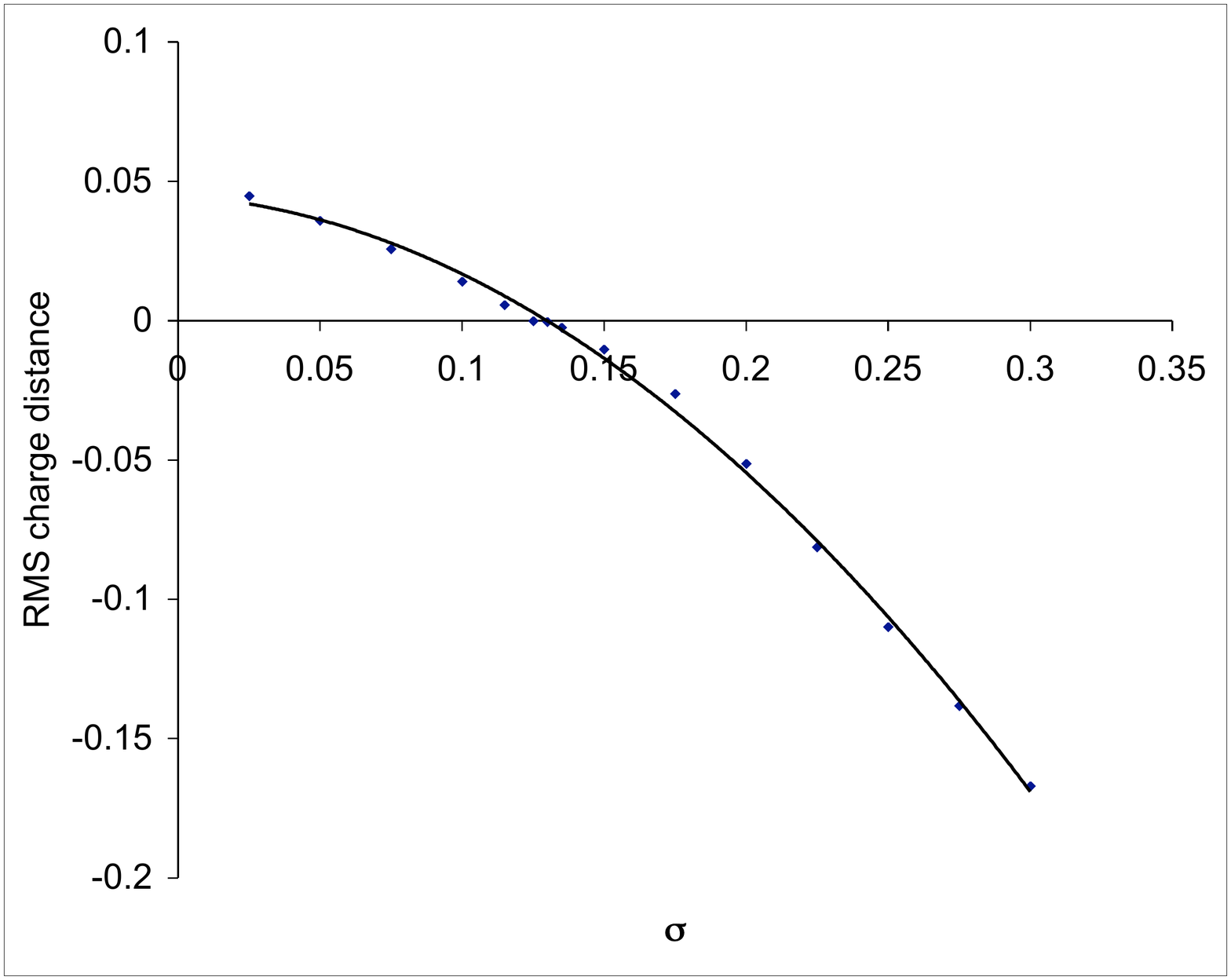}
(b)  \includegraphics[height=5cm,width=6cm,angle=0]%
         {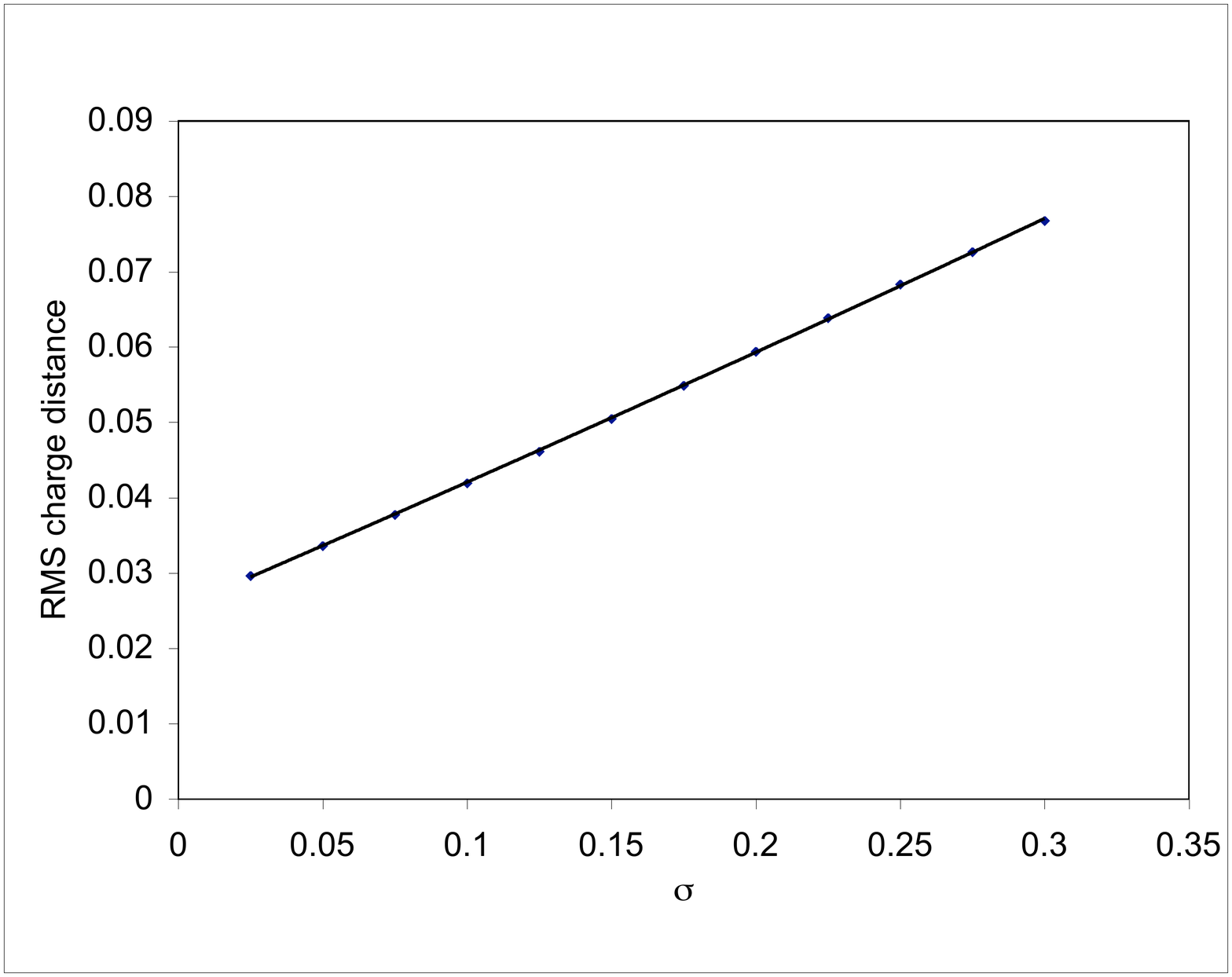}\\
\caption{\label{f:bs1} The root mean squared change of the charge 
within the muffin tins plotted against  $\sigma$ (a) far from the solution and 
(b) near the solution.}
\end{center}
\end{figure}

For our purposes it suffices to give a number of effective controls, with 
reasonable heuristics.  To motivate our strategy we consider the numerical 
experiment, \figr{f:bs1}, showing the root mean squared change of the charge 
within the muffin tins plotted against various values of  $\sigma$ on 
different neighborhoods of the solution.  
Far from the solution, the variation is non-linear 
 (\figr{f:bs1}(a)) over the range $0.05\leq\sigma\leq 0.30$.  Since the step size is directly proportional to the 
size of $\sigma$ \figr{f:bs1}(a) indicates that the linear model is correct only on a small neighborhood
of the current iterate.  If $\sigma$ is too large, the algorithm chooses a step outside the range of validity of 
the local model;  for an LAPW code this will lead to ghost bands and can lead to divergence. As the 
 iterations proceed, that is, on closer neighborhoods to the solution, the change in the charge as a function
of $\sigma$ appears to be linear (\figr{f:bs1}(b)) over the same range, indicating that the 
linear model of the SCF mapping is accurate on a larger neighborhood of the iterate.
Thus we interpret $\sigma$ in the MSBB update as a type of trust-region parameter for a 
linear model of the  SCF mapping.

Our strategy for implementing a dynamic step length $\sigma_n$ has three parts.  
First among these is to constrain $\sigma_n$ so that the 
step in the direction of the unpredicted component has an upper bound that is proportional to the size of the 
predicted component:
\begin{equation}\label{e:nfl}
\sigma_n \leq  R |p_{n}|/|g_{n}|
\end{equation}
where $R$ is a fixed parameter.  
In the context of the MSBB update, this ensures that the unpredicted (and unpredictable) 
component of the step at each iterations does not dominate the total step. 
One has to take some step along this component, as otherwise no new information 
is generated; however if too greedy a step is taken the algorithm can diverge. 
For an ``easy'' problem $R$ can be larger than for a ``difficult'' problem since
for easy problems the unpredicted component of the step is naturally well-scaled.  
Our experience indicates that for robust performance across a wide variety of problems, 
the parameter $R$ is the most important control 
element for $\sigma_n$.
For hard problems a value of $R$ from $0.05$ to $0.15$ works well.
As a second level of control, we bound the total variation between successive scalings:
\begin{equation}\label{e:bound}
\tilde{\sigma}_{n}=\sigma_{n-1}*\max(0.5,\min(2.0,\|g_{n-1}\|/\|g_{n}\|)).
\end{equation}
Note that, by this control, $\sigma_n$ cannot be less than half nor more than twice the previous scaling.  
Note also that we do not reject steps that yield a larger residual $g_n$, but rather reduce the size of the 
step in the unpredicted direction.  In almost all cases a large 
improvement is achieved in the next step by retaining the bad step, though this strategy runs the risk
of developing ghost bands. 
As a third level of control, we include an upper bound on the absolute value of the scaling, $\overline{\sigma}$. 
For our applications we found $\overline{\sigma}\approx 0.1 - 0.2$ to be effective. 
Finally, for the very first cycle we take a small step with
\begin{equation}\label{e:sigma zero}
	\sigma_0  = \overline{\sigma}*(0.1+\exp(-2.0*\max(dQ, dPW/3.5, dRMT))),
\end{equation}
where $dQ$ is the change in the charge within the muffin tins, $dPW$ is the change
in the rescaled plane waves and $dRMT$ is the change of the density 
within the muffin tins. This form is based upon numerical experience with Wien2k, 
and is somewhat conservative.

\subsection{\label{s:summary} Summary}
\begin{algorithm}[Regularized, preconditioned, limited-memory multisecant method]
$~$
\begin{enumerate}
\item[0.] Choose an initial $\rho_0$, $\sigma_0$ according to \eqr{e:sigma zero}, generate
$\rho_1 = \rho_0 + \lambda(F(\rho_0)-\rho_0)$ for $\lambda>0$ some appropriately 
chosen step length (this is the Pratt step \eqr{e:Pratt}), set $n=1$  and fix 
$\alpha>0$ ($10^{-6}$ to $10^{-4}$). 
	\item[1.] If the convergence criterion is met, terminate.  Otherwise, given $S_{n-1}$
and $Y_{n-1}$, whose columns are steps $s_j$ and residual differences $y_j$ 
respectively ($j=n-m,,n-(m-1),\dots,n-1$ for some appropriate number of prior steps, e.g. 
$m=\min\{n,8\}$)  centered on the current point $\rho_n$ as in \eqr{e:Pulay}, 
calculate $A_{n}^{\alpha,\Psi_n,\Omega_n}$ via 
\eqr{e:heur1_regb}  for either MSBB or 
MSGB with the scaling $\Psi_n$ given by \eqr{e:Psin} and the 
preconditioner $\Omega_n$ given by \eqr{e:g^ave}-(\ref{e:precond}).
Determine the value of $\sigma_n$ according to 
           \begin{equation}\label{e:sigma}
           \sigma_n = \min\{\tilde{\sigma}_n, R |p_{n}|/|g_{n}|, \overline{\sigma} \}
           \end{equation}
where $\tilde{\sigma}_n$ is given by \eqr{e:bound} and $\overline{\sigma}$ is some appropriately
chosen upper bound ($0.1$ to $0.2$).
Calculate the next step $\rho_{n+1}$ according to \eqr{e:gen_heur} with $A_n$ replaced by 
$A_{n}^{\alpha,\Psi_n,\Omega_n}$.
        \item[2.] Evaluate $F(\rho_{n+1})$, set $n=n+1$ and repeat Step 1.
\end{enumerate}
\end{algorithm}
\section{\label{s:results} Results}
We test the performance of the algorithm on five examples of increasing physical difficulty,
all run using the Wien2k code \cite{Wien2k} and the PBE functional \cite{PBE}; 
we provide the details below with technical information so they can be reproduced as well as 
reasons for their choice.
\begin{description}
	\item[Model 1] Simple bulk MgO, spin-unpolarized with RMT's of $1.8$ a.u., an RKMAX of $7$ and a 
$5\times 5\times 5$ $k$-point mesh and a Mermin-functional \cite{Mermin65} 
(i.e. Fermi-Dirac distribution) 
with a temperature of $0.0068$eV. This is a very easy to solve problem.
	\item[Model 2] Bulk Pd, spin-unpolarized with RMT's of $2.0$ a.u., an RKMAX of $7.5$, a 
$5\times 5\times 5$ $k$-point mesh and a Mermin-functional with a temperature of $0.0068$eV. 
This is slightly harder because of the possibility of sloshing between the $d$-electron states and the fact 
that one should use a larger sampling of reciprocal space.
	\item[Model 3] A bulk silicon cell with an RMT of $2.16$ a.u., an RKMAX of $7.0$, a 
$6\times 6\times 6$ $k$-point mesh and a Mermin-functional with a temperature of $0.0013$eV.
	\item[Model 4] A $2\times 2\times 2$ Pd supercell with a vacancy at the origin, RMT's of $2.5$ a.u., 
an RKMAX of $6.5$, a $k$-point mesh of $3\times 3\times 3$ and a Mermin-functional with a 
temperature of $0.0068eV$. Here, in addition to sloshing between $d$-electron states one can have 
longer-range dielectric sloshing. In addition, this is a poorly constructed problem because the RKMAX is 
too small as is the $k$-point mesh.	
	\item[Model 5] A  $4.757\times 4.757\times 34.957$ a.u., spin-polarized (111) fcc nickel surface 
with seven atoms in the range $-1/3 \leq z \leq 1/3$. Technical parameters were RMTs of $2.13$, an 
RKMAX of $7$ and a $11\times 11\times 1$ $k$-point mesh, also with a Mermin-function 
temperature of $0.0068eV$. It should be noted that the two surfaces are sufficiently close together, so 
there is real electron density in the vacuum. In this case one can have spin sloshing, $d$-electron 
sloshing as well as long-range Coulomb sloshing of electrons in the vacuum.
\end{description}

\begin{figure}
\begin{center}
  \includegraphics[height=5.5cm,width=8cm,angle=0]%
         {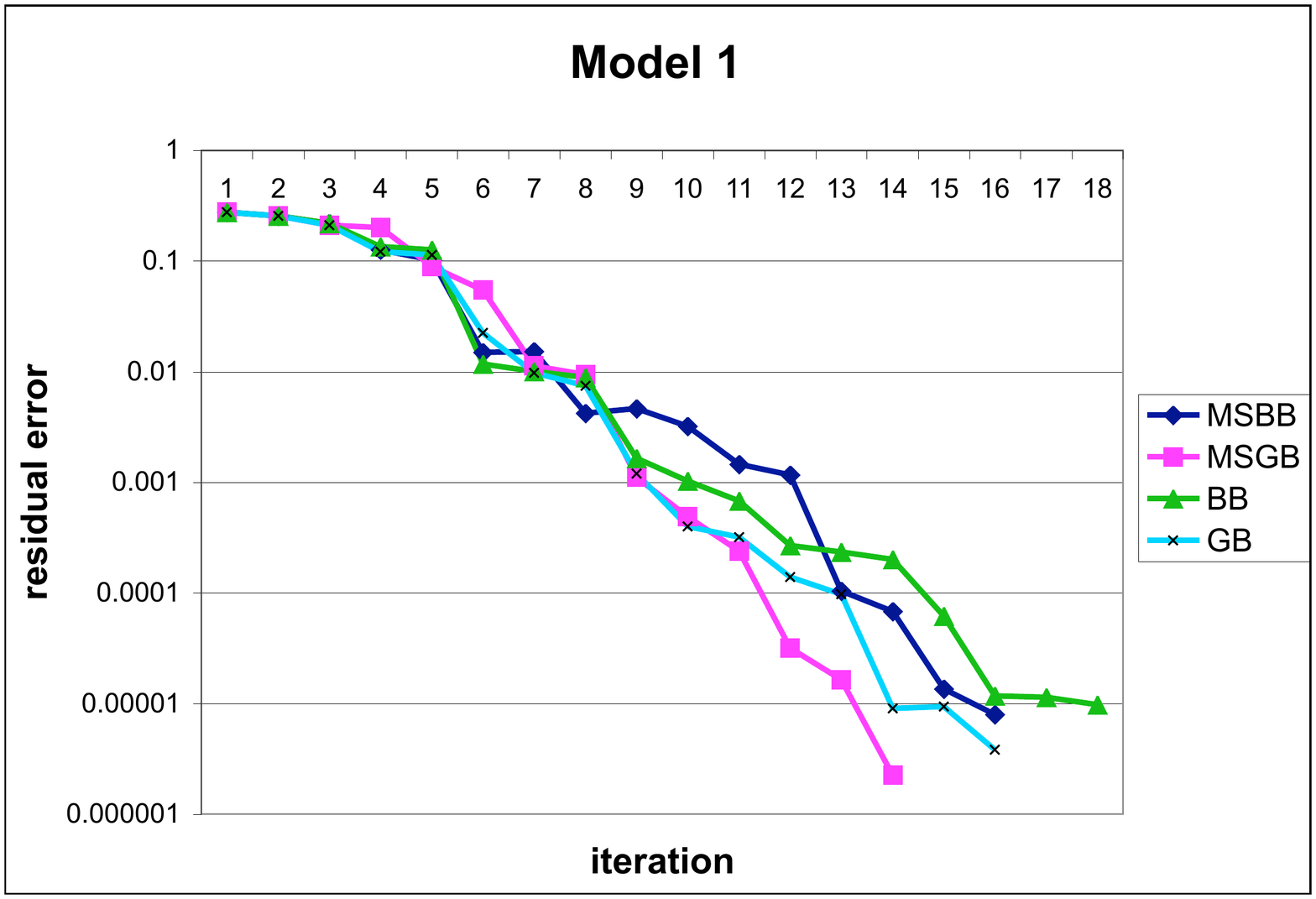}
\hfill
  \includegraphics[height=5.5cm,width=8cm,angle=0]%
         {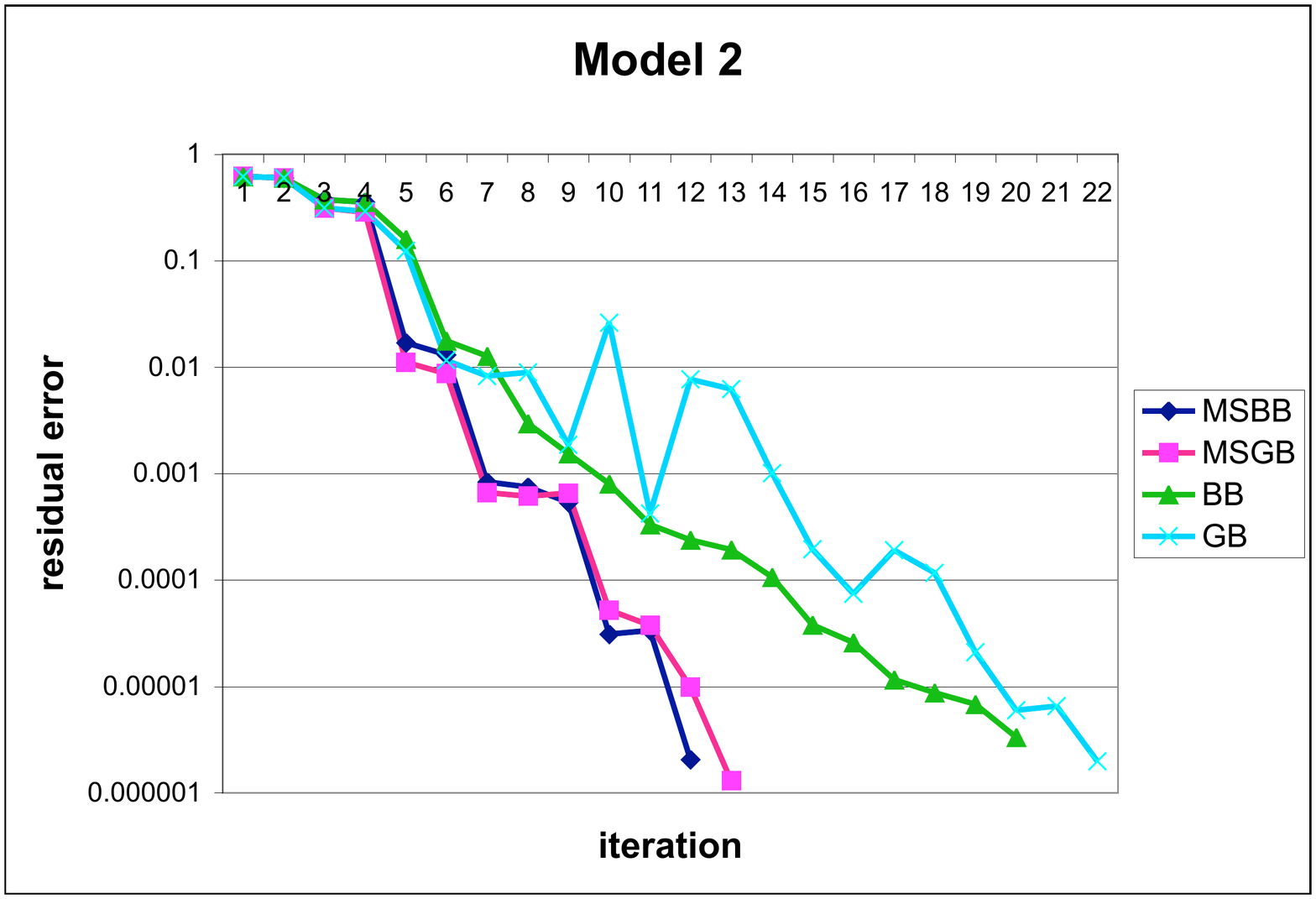}\\
  \includegraphics[height=5.5cm,width=8cm,angle=0]%
         {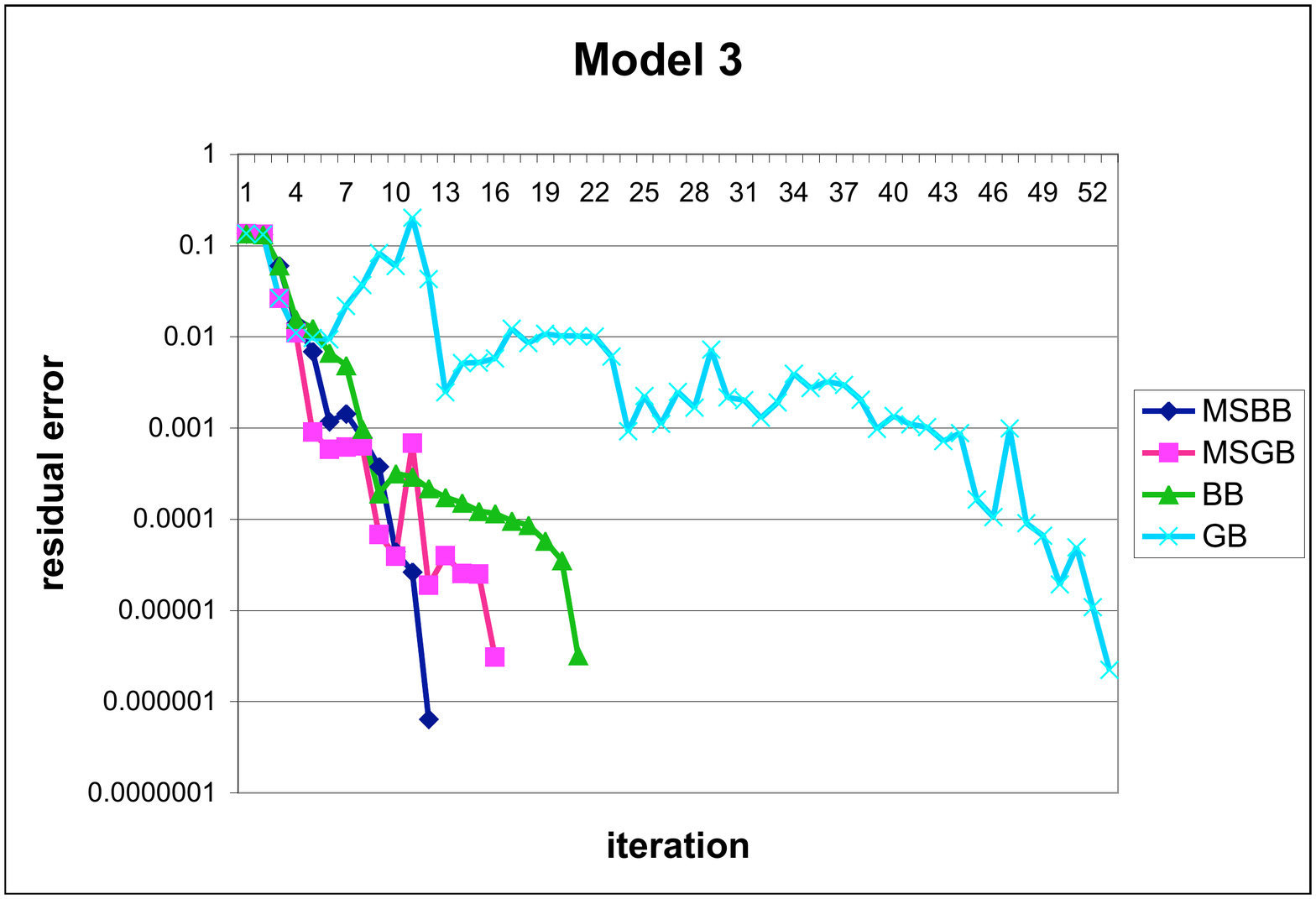}
\hfill
  \includegraphics[height=5.5cm,width=8cm,angle=0]%
         {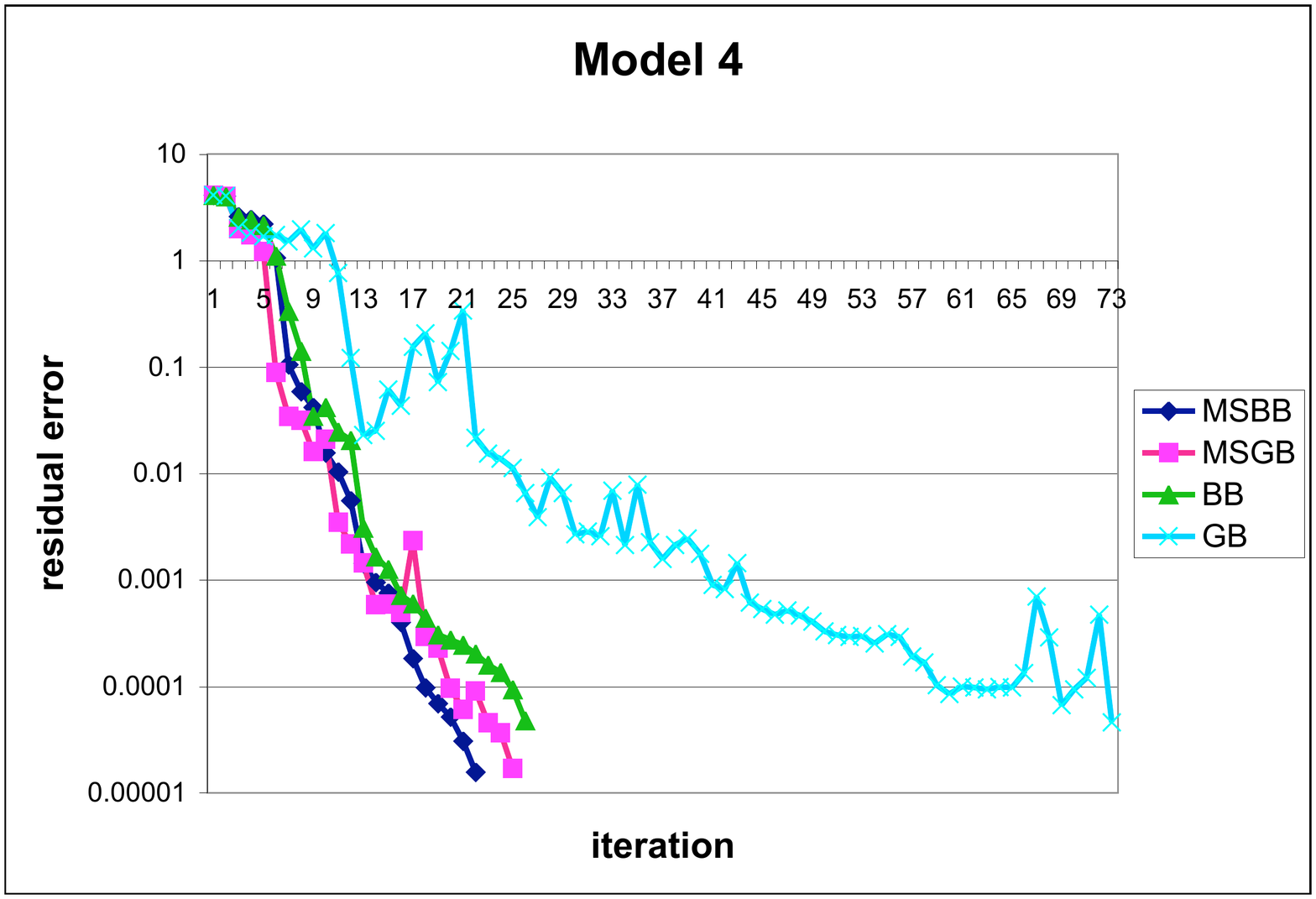}\\
  \includegraphics[height=5.5cm,width=8cm,angle=0]%
         {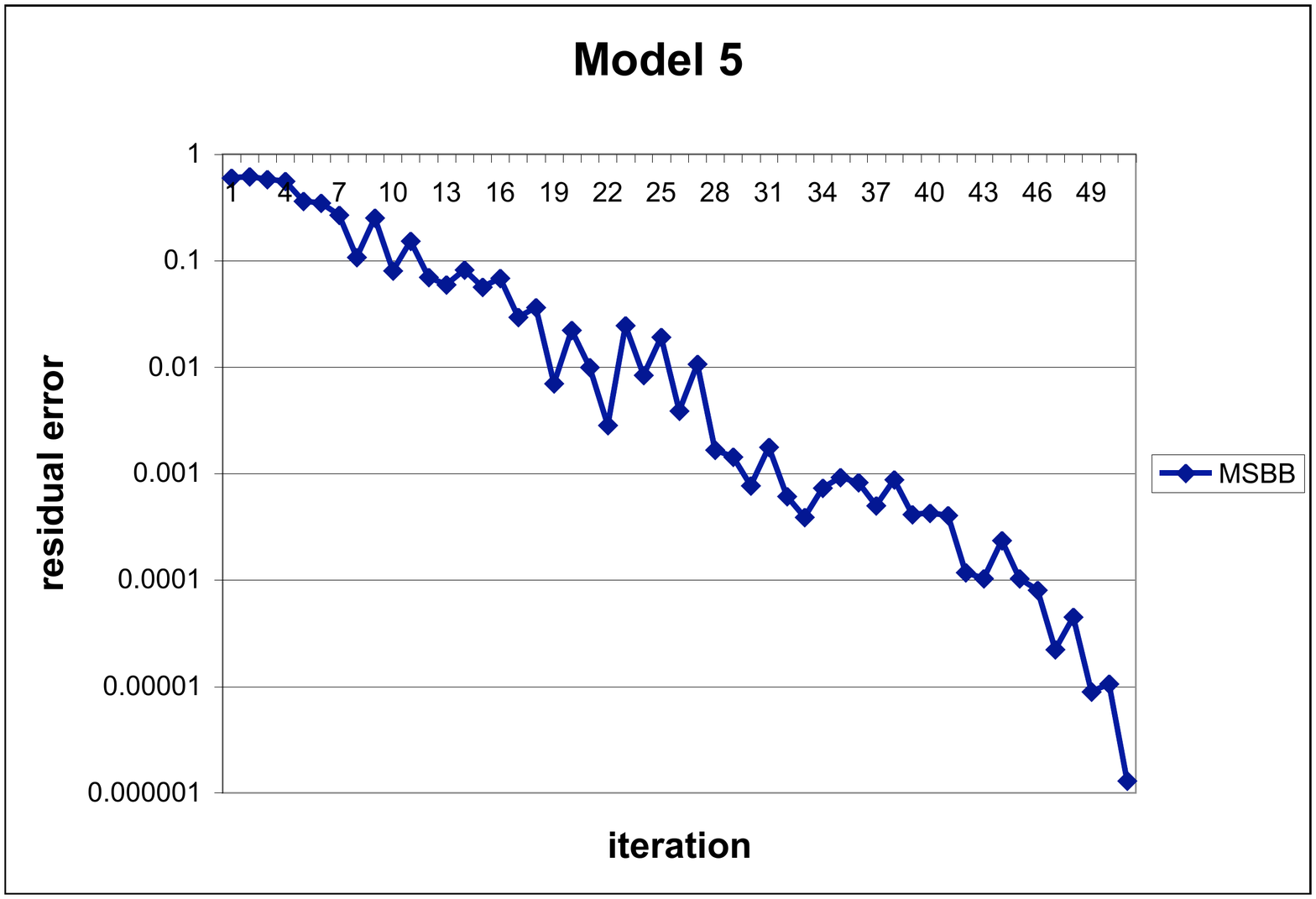}
\caption{\label{f:Convergence}Plot of the convergence for models 1-5 
using the multisecant update based on Broyden's first method (MSGB) and second 
method (MSBB), compared to Broyden's second 
method (BB) and Broyden's
first method (GB).  In model 5 the only algorithm to converge is MSBB.}
\end{center}
\end{figure}

In all cases we started from densities calculated as a sum of independent atoms, and the 
calculations were run with both forms of Broyden multisecants given by \eqr{e:MSBB} 
and \eqr{e:MSGB}, as well as the 
more conventional Broyden first \eqr{e:GB} and second \eqr{e:BB} methods. 
Convergence criteria were an energy change of $10^{-5}$ Rydbergs and an RMS convergence 
of the charge within the muffin 
tins of $10^{-5}$ electrons.
For the multisecant implementations eight prior memory steps were used.
To simplify the results, unless noted otherwise we used fixed values of the regularization 
parameter $\alpha$ of $10^{-4}$ 
and $R=0.1$.  In almost all cases, \figr{f:Convergence} shows that the convergence appears to be 
linear, although the precision of the calculations does not allow one to observe the final 
asymptotic behavior, including rates of convergence, of the algorithms.

\begin{table}[ht]
\caption{\label{table1} {\small \noindent Iterations to convergence as a function of $\sigma$ for 
models $1-5$ with fixed $\alpha=10^{-4}$ and $R=0.1$. The mean and standard deviation are for 
$\overline{\sigma}$ between $0.05$ and $0.8$ for Models $1-3$, $0.05$ to $0.5$ for Models $4$ and $5$.}}
\begin{center}
\begin{tabular}{|c||c|c||c|c||c|c||c|c|}\hline
& \multicolumn{2}{c||}{$~$}& \multicolumn{2}{c||}{$~$}& \multicolumn{2}{c||}{$~$}&
 \multicolumn{2}{c|}{$~$}\\
& \multicolumn{2}{c||}{{\large MSBB}}& \multicolumn{2}{c||}{{\large MSGB}}&
 \multicolumn{2}{c||}{{\large BB}}& \multicolumn{2}{c|}{{\large GB}}\\
& mean & stdev & mean & stdev & mean & stdev & mean & stdev \\
\hhline{|~||=|=||=|=||=|=||=|=|}
Model 1 & 16.22 & 0.44 & 14.56  & 1.01  & 18.67 & 1.66 & 22.44 & 12.71 \\
\hhline{-||-|-||-|-||-|-||-|-|}
Model 2  & 12 & 0 & 12.89 & 0.33 & 20.11 & 0.78 & 39.78 & 9.58 \\
\hhline{-||-|-||-|-||-|-||-|-|}
Model 3 & 15.44 & 2.35 & 16.78 & 0.67 & 25.22 & 2.95 & 57.56 & 7.76\\
\hhline{-||-|-||-|-||-|-||-|-|}
Model 4 & 24.17 & 2.04 & 29.17 & 4.92 & -- & -- & -- & --\\
\hhline{-||-|-||-|-||-|-||-|-|}
Model 5 & 54.60 & 3.51 & -- & -- & -- & -- & -- & -- \\
\hline
\end{tabular}
\end{center}
\end{table}

\begin{figure}
\begin{center}
  \includegraphics[height=7cm,width=8cm,angle=0]%
         {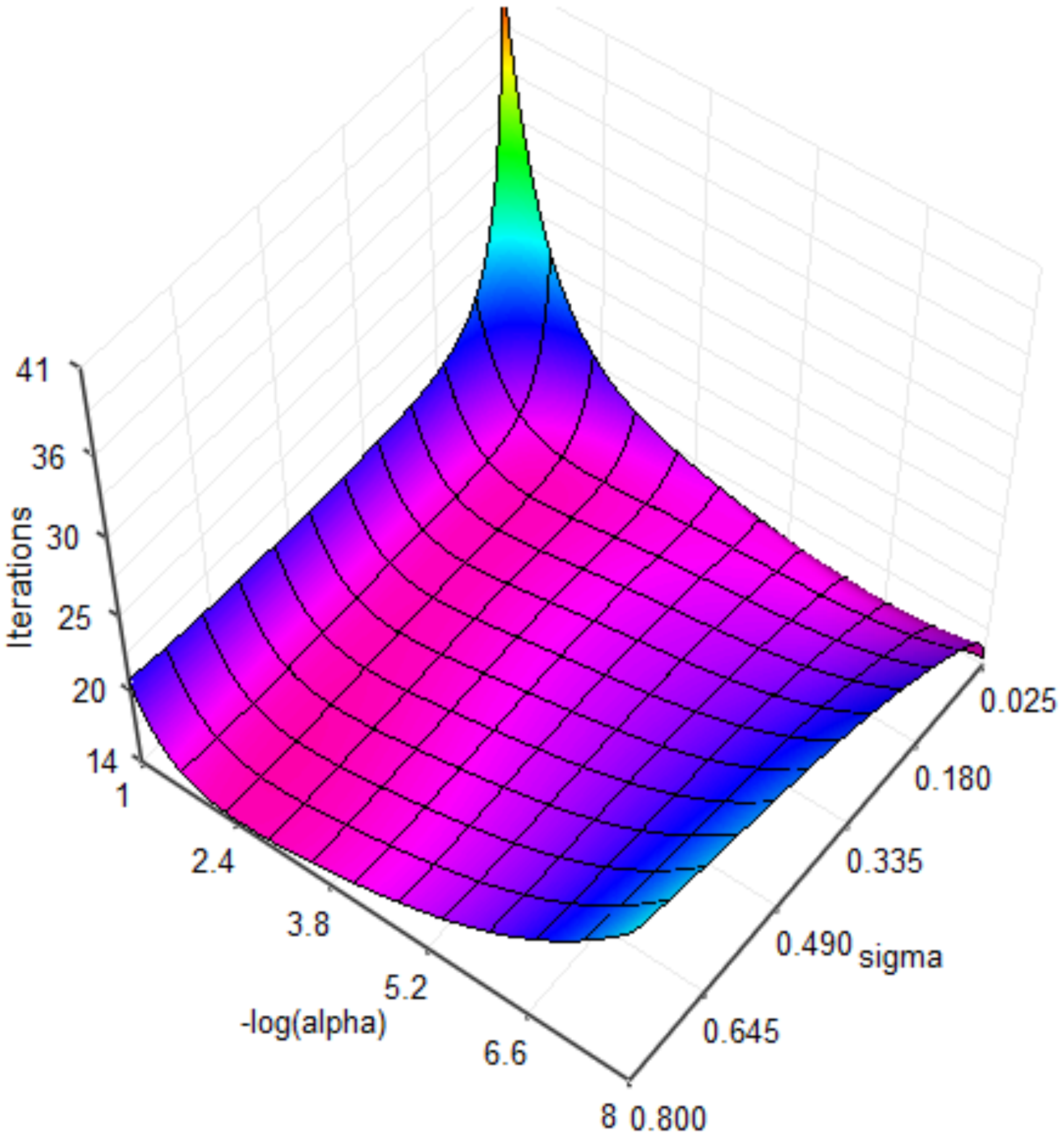}
\caption{\label{f:bs3} Number of iterations to convergence for MSGB on Model 1 as a function
of the value of the algorithm parameters $\alpha$ and $\sigma$.  The surface plot has 
been fitted by a polynomial to obtain smooth contours representing the behavior.}
\end{center}
\end{figure}

For the very simple Model $1$ all the methods converge quickly and the parameter $\overline{\sigma}$ has no
significant impact on performance.  The MSGB
method is slightly faster, but as the latter results indicate this is an exception. 
If $\overline{\sigma}$ is too small (below $~0.025$) convergence is slower  
as illustrated in the plot of the number of iterations to convergence versus both $\overline{\sigma}$ and 
$\alpha$ shown in \figr{f:bs3}. 
Interestingly, even for this very simple case the multisecant methods are significantly faster. 

For the slightly more complicated Model 2, both multisecant methods converge rapidly, whereas the 
BB method converges more slowly and the GB method is worst by a significant margin. 
The principal difference between Models 1 and 2 is that in Model 1 there are large changes 
during the iterations both within the muffin-tins, as well as for the plane waves, whereas in Model 2 
almost all the changes are in the plane waves.  This supports the rule-of-thumb discussed earlier
that one should make the muffin tins as large as possible without overlapping.

With Model 3 the multisecant methods significantly outperform the classical secant methods. 
For bulk silicon much of the covalent bonding lies in the interstitial region.  We conjecture, therefore,
that the improvement is due to the improved step direction and size for the multisecant methods that
allow these methods to handle the greater variations of the Kohn -Sham mapping for this basis set.
 
The same trend continues with both Model 4 and Model 5 to the extent that BB and GB only converge for 
``good'' values of $\overline{\sigma}$ (which have to be found by trial and error) and in many cases diverge. 
For the hardest problem we report here, Model 5, only the MSBB method converged. If one
added a line search the other methods would probably converge albeit less rapidly and with a many more 
SCF evaluations.

The $\sigma$ parameter in the MSBB update gives one direct control over the size of the steps, which is
an important feature for models 
with strong variations.  The control of steps is less immediate for the MSGB update and involves a more 
sensitive coupling of the regularization parameter $\alpha$ and the step size parameter $\sigma$.  This is 
illustrated by the 
greater variance in performance of the MSGB update versus MSBB for models $1$, $2$, $4$, and $5$ shown 
in Table \ref{table1}. 

\section{Discussion}
To summarize the main points of this work:
\begin{itemize}
	\item We argue that for DFT problems, where many physically interesting models result in noncontractive
SCF mappings, one should consider the information from previous points of
	the SCF cycle more as samples of a higher-dimensional space than as part of a deterministic path. As a 
	consequence multisecant methods are better than sequential secant updates, as born out in the results.
	\item There is a fundamental difference between methods based upon Broyden's first (GB) and second 
(BB) methods in terms of the space they operate in. The second method is more robust
	and handles poorly constructed, (nearly) ill-posed problems better  -- in general these are the 
more interesting physical problems.
	\item Scaling, regularization and preconditioning have a significant impact on algorithm performance.
Moreover, regularization acts simultaneously to reduce instabilities due both to linear
	dependencies as well as to deficiencies in the model.  
	\item Controlling the step size $\sigma_n$ along the direction about which no information is available 
is critical.  For difficult problems, this step should in general be {\em smaller} than for easy problems.
	\item The multisecant method based upon Broyden's second formulation (MSBB) with appropriate safeguards
	 simply and quickly solves problems which may defeat a novice, sometimes even an expert.
\end{itemize}
The method we have detailed (MSBB) is robust and has been part of the main Wien2k distribution (7.3) 
for some months without any apparent problems. Even in the hands of an experienced user for
complicated problems such as LDA+U we have been told of cases where the MSBB version is three times faster
than the earlier BB code. The default values of $\alpha=10^{-4}$ and $R=0.1$ will  
be approximately correct for a pseudopotential code where 
preconditioning the variables is not necessary though there are strong variations Kohn-Sham mapping.
We have not attempted to impliment the MSBB algorithm for a pseudopotential code 
but see no reason why it should not work at least as well.
One can of course adjust these parameters to improve 
a single problem, but we recommend values that
perhaps are slightly slower in a few cases, but more robust for a wide variety of problems.
There may also be ways to stabilize MSGB so that it could
possibly work better for pseudopotential codes where preconditioning is easier.

Some additional comments are appropriate about the role of the term in the regularization. As mentioned 
earlier, we are using this \textit{simultaneously} in three ways, firstly 
as a standard regularization technique to avoid 
ill-conditioning associated with near linear dependence of the columns of  
$Y_n$, secondly as a Levenberg-Marquardt-type strategy to rotate the step and thirdly 
in a standard Wiener filter sense to account for model uncertainty.  
The regularization parameter can be considered to scale proportionally to the noise or uncertainty 
in the secant equations. Far from the solution the quasi-Newton step may not be appropriate, 
suggesting that one should use a larger regularization. Similarly, near the solution if the quasi-Newton 
step is accurate, it will yield faster rates of convergence, in which case one would choose a smaller
regularization.  While one could dynamically adjust the regularization parameter, for 
our numerical experiments we choose a relatively large fixed value of $\alpha$ ( $10^{-4}$).   
This, in our experience, yields adequate overall convergence and better convergence 
in the ``dangerous'' early stages of the iterations.

We emphasize once again the link between convergence of the mixing process and the functional
properties of the underlying Kohn-Sham mapping.  A poorly constructed problem 
will in most cases converge much more slowly than a well constructed one.  This may be a consequence of 
short-cuts in the DFT calculation, e.g. too few $k$-points or numerical errors in an iterative 
diagonalization, or it can be due to a poorly constructed Hamiltonian or perhaps density functional. 
For the general user poor convergence should be taken as a suggestion that the model of the physics 
may not have been properly constructed.

The fact that we that we obtain improvement under the assumption that
most models of physical interest do not lead to contractive, or more
generally monotone SCF mappings raises some questions. It is well
established that current density functionals are inexact descriptions
of the physics, but the exact analytic properties of many physical
systems are unknown.  In particular, for many systems it is not known
whether the SCF operator is monotone, let alone that it has fixed
points, although it is hard to conceive of an experimentally observable
equilibrium structure that does not have fixed pionts.  
An interesting question to raise is whether the SCF operator is monotonic 
with the "true" density functional that correctly describes the physics.
Since in many cases the effective potential $V_\rho$ has no closed form, 
it is not known
whether many of these theoretical properties are verifiable.  It is
tempting to infer analytic properties from numerical experiments --
and we have made numerical progress by doing just this -- but one
cannot on numerical evidence alone determine the extent to which
numerical behavior is indicative of the true nature of the physical
system.  As a final speculation, we raise the question of whether the
character of the SCF mapping can be experimentally measured, or
whether this type of behavior is a mathematical anomaly resulting 
from being much further away from equilibrium than any feasible 
experimental system will ever be.

There are several directions of research with regard to algorithms.
Firstly, the heuristics for adjusting the step size $\sigma_n$ need to be put on firm mathematical
footing.  This would accompany a study of the asymptotic behavior of the algorithm and is 
the subject of future research.  While the analysis of \eqr{e:heur1} has attractive interpretations 
in terms of nearest points in 
the range and space orthogonal to the prior data, the notion of ``nearest'' is with respect to the 
usual Euclidean ($L^2$) norm, which is biased towards outliers.  One could consider the 
development of algorithms based on weighted norms, or even non-Euclidean prox mappings as 
opposed to those detailed in Subsection \ref{s:prox}.  The $\Omega_n$ considered by 
\cite{VanderbiltLouie84, Johnson88, CrittinBierlaire03} is in the spirit of weighted norms.  
Other areas for improvement could be found in the initialization 
of the iterations.  We used the Pratt step, however one could use information from a previous 
SCF iteration.   

Finally, while we have used some physics in
helping to design the algorithm, there may be more that could be exploited.  We
find particularly appealing the observation 
discussed at the beginning of Section \ref{s:new stuff} that the density appears to 
be separable into distinct subsets.  
One might envision tailoring algorithms to exploit this property.  For instance, one
could iterate on the components of the density associated with the muffin-tins, while 
holding the interstitial electron density fixed.  
Alternatively, one could 
iterate on the $sp$-electron density holding the $d$-electron density fixed,
or one could iterate on other observables such as the spin associated 
with a particular atom.  Such an approach might allow one to isolate irregular
variables within the SCF mapping and design algorithms accordingly.  This general
approach is known as {\em operator splitting} about which there is a vast literature.
(see, for instance \cite{Luke07} and references therein).  
This would allow one to isolate the analytical properties of the SCF operator and 
work more directly with specific physical quantities.

\begin{acknowledgments}
This work was funded by NSF under Grants \#DMR-0455371/001 (LDM) and
\#DMS-0712796 (DRL).
\end{acknowledgments}
%
%
%

\begin{thebibliography}{59}
\expandafter\ifx\csname natexlab\endcsname\relax\def\natexlab#1{#1}\fi
\expandafter\ifx\csname bibnamefont\endcsname\relax
  \def\bibnamefont#1{#1}\fi
\expandafter\ifx\csname bibfnamefont\endcsname\relax
  \def\bibfnamefont#1{#1}\fi
\expandafter\ifx\csname citenamefont\endcsname\relax
  \def\citenamefont#1{#1}\fi
\expandafter\ifx\csname url\endcsname\relax
  \def\url#1{\texttt{#1}}\fi
\expandafter\ifx\csname urlprefix\endcsname\relax\def\urlprefix{URL }\fi
\providecommand{\bibinfo}[2]{#2}
\providecommand{\eprint}[2][]{\url{#2}}

\bibitem[{\citenamefont{Kohn and J.}(1965)}]{KohnSham65}
\bibinfo{author}{\bibfnamefont{W.}~\bibnamefont{Kohn}} \bibnamefont{and}
  \bibinfo{author}{\bibfnamefont{S.~L.} \bibnamefont{J.}},
  \bibinfo{journal}{Phys. Rev.} \textbf{\bibinfo{volume}{140}},
  \bibinfo{pages}{A 1133} (\bibinfo{year}{1965}).

\bibitem[{\citenamefont{Mermin}(1965)}]{Mermin65}
\bibinfo{author}{\bibfnamefont{N.~D.} \bibnamefont{Mermin}},
  \bibinfo{journal}{Phys. Rev.} \textbf{\bibinfo{volume}{137}},
  \bibinfo{pages}{A 1441} (\bibinfo{year}{1965}).

\bibitem[{\citenamefont{Prodan}(2005)}]{Prodan05}
\bibinfo{author}{\bibfnamefont{E.}~\bibnamefont{Prodan}}, \bibinfo{journal}{J.
  Phys. A.} \textbf{\bibinfo{volume}{38}}, \bibinfo{pages}{5647}
  (\bibinfo{year}{2005}).

\bibitem[{\citenamefont{Blaha et~al.}(2006)\citenamefont{Blaha, Schwarz,
  Madsen, Kvasnicka, and Luitz}}]{Wien2k}
\bibinfo{author}{\bibfnamefont{P.}~\bibnamefont{Blaha}},
  \bibinfo{author}{\bibfnamefont{K.}~\bibnamefont{Schwarz}},
  \bibinfo{author}{\bibfnamefont{G.}~\bibnamefont{Madsen}},
  \bibinfo{author}{\bibfnamefont{D.}~\bibnamefont{Kvasnicka}},
  \bibnamefont{and} \bibinfo{author}{\bibfnamefont{J.}~\bibnamefont{Luitz}},
  \emph{\bibinfo{title}{{WIEN2k}, An Augmented Plane Wave + Local Orbitals
  Program for Calculating Crystal Properties}} (\bibinfo{publisher}{Institute
  for Materials Chemistry, TU Vienna}, \bibinfo{address}{{\ttfamily
  http://www.wien2k.at/}}, \bibinfo{year}{2006}).

\bibitem[{\citenamefont{Crittin and Bierlaire}(2003)}]{CrittinBierlaire03}
\bibinfo{author}{\bibfnamefont{F.}~\bibnamefont{Crittin}} \bibnamefont{and}
  \bibinfo{author}{\bibfnamefont{M.}~\bibnamefont{Bierlaire}}, in
  \emph{\bibinfo{booktitle}{{Proceedings of the 3rd Swiss Transport Research
  Conference}}} (\bibinfo{organization}{STRC}, \bibinfo{year}{2003}).

\bibitem[{\citenamefont{Johnson}(1988)}]{Johnson88}
\bibinfo{author}{\bibfnamefont{D.~D.} \bibnamefont{Johnson}},
  \bibinfo{journal}{Phys. Rev. B} \textbf{\bibinfo{volume}{38}},
  \bibinfo{pages}{12807} (\bibinfo{year}{1988}).

\bibitem[{\citenamefont{Kawata et~al.}(1998)\citenamefont{Kawata, Cortis, and
  Friesner}}]{Kawata98}
\bibinfo{author}{\bibfnamefont{M.}~\bibnamefont{Kawata}},
  \bibinfo{author}{\bibfnamefont{C.~M.} \bibnamefont{Cortis}},
  \bibnamefont{and} \bibinfo{author}{\bibfnamefont{R.~A.}
  \bibnamefont{Friesner}}, \bibinfo{journal}{J. Chem. Phys.}
  \textbf{\bibinfo{volume}{108}}, \bibinfo{pages}{4426} (\bibinfo{year}{1998}).

\bibitem[{\citenamefont{Pulay}(1980)}]{Pulay80}
\bibinfo{author}{\bibfnamefont{P.}~\bibnamefont{Pulay}},
  \bibinfo{journal}{Chem. Phys. Lett.} \textbf{\bibinfo{volume}{73}},
  \bibinfo{pages}{393} (\bibinfo{year}{1980}).

\bibitem[{\citenamefont{Srivastava}(1984)}]{Srivastava}
\bibinfo{author}{\bibfnamefont{G.~P.} \bibnamefont{Srivastava}},
  \bibinfo{journal}{J. Phys. A: Math. Gen.} \textbf{\bibinfo{volume}{17}},
  \bibinfo{pages}{L317} (\bibinfo{year}{1984}).

\bibitem[{\citenamefont{Vanderbilt and Louie}(1984)}]{VanderbiltLouie84}
\bibinfo{author}{\bibfnamefont{D.}~\bibnamefont{Vanderbilt}} \bibnamefont{and}
  \bibinfo{author}{\bibfnamefont{S.~G.} \bibnamefont{Louie}},
  \bibinfo{journal}{Phys. Rev. B} \textbf{\bibinfo{volume}{30}},
  \bibinfo{pages}{6118} (\bibinfo{year}{1984}).

\bibitem[{\citenamefont{Singh et~al.}(1986)\citenamefont{Singh, Krakauer, and
  Wang}}]{Singh86}
\bibinfo{author}{\bibfnamefont{D.}~\bibnamefont{Singh}},
  \bibinfo{author}{\bibfnamefont{H.}~\bibnamefont{Krakauer}}, \bibnamefont{and}
  \bibinfo{author}{\bibfnamefont{C.~S.} \bibnamefont{Wang}},
  \bibinfo{journal}{Phys. Rev. B} \textbf{\bibinfo{volume}{34}},
  \bibinfo{pages}{8391 } (\bibinfo{year}{1986}).

\bibitem[{\citenamefont{Moreau}(1962)}]{Moreau62}
\bibinfo{author}{\bibfnamefont{J.~J.} \bibnamefont{Moreau}},
  \bibinfo{journal}{Comptes Rendus de l'Acad\'emie des Sciences de Paris}
  \textbf{\bibinfo{volume}{255}}, \bibinfo{pages}{2897} (\bibinfo{year}{1962}).

\bibitem[{\citenamefont{Boyd and Vandenberghe}(2003)}]{BoydVan03}
\bibinfo{author}{\bibfnamefont{S.}~\bibnamefont{Boyd}} \bibnamefont{and}
  \bibinfo{author}{\bibfnamefont{L.}~\bibnamefont{Vandenberghe}},
  \emph{\bibinfo{title}{Convex Optimization}} (\bibinfo{publisher}{Oxford
  University Press}, \bibinfo{address}{New York}, \bibinfo{year}{2003}).

\bibitem[{\citenamefont{Broyden}(1965)}]{Broyden65}
\bibinfo{author}{\bibfnamefont{C.~G.} \bibnamefont{Broyden}},
  \bibinfo{journal}{Mathematics of Computation} \textbf{\bibinfo{volume}{19}},
  \bibinfo{pages}{577} (\bibinfo{year}{1965}).

\bibitem[{\citenamefont{Dennis and Schnabel}(1996)}]{DennisSchnabel96}
\bibinfo{author}{\bibfnamefont{J.~E.} \bibnamefont{Dennis}} \bibnamefont{and}
  \bibinfo{author}{\bibfnamefont{R.}~\bibnamefont{Schnabel}},
  \emph{\bibinfo{title}{Numerical Methods for Unconstrained Optimization and
  Nonlinear Equations}} (\bibinfo{publisher}{Prentice Hall},
  \bibinfo{address}{Englewood Cliffs, NJ}, \bibinfo{year}{1996}).

\bibitem[{\citenamefont{Dennis and Mor\'e}(1977)}]{DennisMore77}
\bibinfo{author}{\bibfnamefont{J.~E.} \bibnamefont{Dennis}} \bibnamefont{and}
  \bibinfo{author}{\bibfnamefont{J.~J.} \bibnamefont{Mor\'e}},
  \bibinfo{journal}{SIAM Rev.} \textbf{\bibinfo{volume}{19}},
  \bibinfo{pages}{46} (\bibinfo{year}{1977}).

\bibitem[{\citenamefont{Barnes}(1965)}]{Barnes65}
\bibinfo{author}{\bibfnamefont{J.~G.~P.} \bibnamefont{Barnes}},
  \bibinfo{journal}{Comput. J.} \textbf{\bibinfo{volume}{8}},
  \bibinfo{pages}{66} (\bibinfo{year}{1965}).

\bibitem[{\citenamefont{Gay and Schnabel}(1978)}]{GaySchnabel78}
\bibinfo{author}{\bibfnamefont{D.~M.} \bibnamefont{Gay}} \bibnamefont{and}
  \bibinfo{author}{\bibfnamefont{R.~B.} \bibnamefont{Schnabel}}, in
  \emph{\bibinfo{booktitle}{Nonlinear Programming}}, edited by
  \bibinfo{editor}{\bibfnamefont{O.~L.} \bibnamefont{Mangasarian}},
  \bibinfo{editor}{\bibfnamefont{R.~R.} \bibnamefont{Meyer}}, \bibnamefont{and}
  \bibinfo{editor}{\bibfnamefont{S.~M.} \bibnamefont{Robinson}}
  (\bibinfo{publisher}{Academic Press}, \bibinfo{address}{New York},
  \bibinfo{year}{1978}), vol.~\bibinfo{volume}{3}, pp.
  \bibinfo{pages}{245--281}.

\bibitem[{\citenamefont{Mart\'inez and Zambaldi}(1992)}]{Martinez92c}
\bibinfo{author}{\bibfnamefont{J.~M.} \bibnamefont{Mart\'inez}}
  \bibnamefont{and} \bibinfo{author}{\bibfnamefont{M.~C.}
  \bibnamefont{Zambaldi}}, \bibinfo{journal}{Optim. Method Software}
  \textbf{\bibinfo{volume}{1}}, \bibinfo{pages}{129} (\bibinfo{year}{1992}).

\bibitem[{\citenamefont{Spedicato and Huang}(1997)}]{Spedicato97}
\bibinfo{author}{\bibfnamefont{E.}~\bibnamefont{Spedicato}} \bibnamefont{and}
  \bibinfo{author}{\bibfnamefont{Z.}~\bibnamefont{Huang}},
  \bibinfo{journal}{Comput.} \textbf{\bibinfo{volume}{58}}, \bibinfo{pages}{69}
  (\bibinfo{year}{1997}).

\bibitem[{\citenamefont{Luksan and Vlcek}(1998)}]{LucksanVlcek98}
\bibinfo{author}{\bibfnamefont{L.}~\bibnamefont{Luksan}} \bibnamefont{and}
  \bibinfo{author}{\bibfnamefont{J.}~\bibnamefont{Vlcek}},
  \bibinfo{journal}{Optim. Methods Software} \textbf{\bibinfo{volume}{8}},
  \bibinfo{pages}{185} (\bibinfo{year}{1998}).

\bibitem[{\citenamefont{Mart\'inez}(2000)}]{Martinez00}
\bibinfo{author}{\bibfnamefont{J.~M.} \bibnamefont{Mart\'inez}},
  \bibinfo{journal}{J. Comp. Appl. Math.} \textbf{\bibinfo{volume}{124}},
  \bibinfo{pages}{97} (\bibinfo{year}{2000}).

\bibitem[{\citenamefont{IP and Todd}(1988)}]{IPTodd88}
\bibinfo{author}{\bibfnamefont{C.~M.} \bibnamefont{IP}} \bibnamefont{and}
  \bibinfo{author}{\bibfnamefont{M.~J.} \bibnamefont{Todd}},
  \bibinfo{journal}{SIAM J. Numer. Anal.} \textbf{\bibinfo{volume}{25}},
  \bibinfo{pages}{206} (\bibinfo{year}{1988}).

\bibitem[{\citenamefont{Byrd et~al.}(1994)\citenamefont{Byrd, Nocedal, and
  Schnabel}}]{ByrdNocedalSchnabel94}
\bibinfo{author}{\bibfnamefont{R.~H.} \bibnamefont{Byrd}},
  \bibinfo{author}{\bibfnamefont{J.}~\bibnamefont{Nocedal}}, \bibnamefont{and}
  \bibinfo{author}{\bibfnamefont{R.~B.} \bibnamefont{Schnabel}},
  \bibinfo{journal}{Math. Prog.} \textbf{\bibinfo{volume}{63}},
  \bibinfo{pages}{129} (\bibinfo{year}{1994}).

\bibitem[{\citenamefont{Davidon}(1959)}]{Davidon59}
\bibinfo{author}{\bibfnamefont{W.~C.} \bibnamefont{Davidon}},
  \bibinfo{type}{Tech. Rep.}, \bibinfo{institution}{Atomic Energy Commission
  Research and Development Report AWL--5990}, \bibinfo{address}{Argonne
  National Laboratory, Argonne, IL} (\bibinfo{year}{1959}).

\bibitem[{\citenamefont{Davidon}(1975)}]{Davidon75}
\bibinfo{author}{\bibfnamefont{W.~C.} \bibnamefont{Davidon}},
  \bibinfo{journal}{Math. Programming} \textbf{\bibinfo{volume}{9}},
  \bibinfo{pages}{1} (\bibinfo{year}{1975}).

\bibitem[{\citenamefont{Oren and Spedicato}(1976)}]{Oren}
\bibinfo{author}{\bibfnamefont{S.~S.} \bibnamefont{Oren}} \bibnamefont{and}
  \bibinfo{author}{\bibfnamefont{E.}~\bibnamefont{Spedicato}},
  \bibinfo{journal}{Math.Prog.} \textbf{\bibinfo{volume}{10}},
  \bibinfo{pages}{70} (\bibinfo{year}{1976}).

\bibitem[{\citenamefont{Shanno and Phua}(1978)}]{Phua}
\bibinfo{author}{\bibfnamefont{D.~F.} \bibnamefont{Shanno}} \bibnamefont{and}
  \bibinfo{author}{\bibfnamefont{K.}~\bibnamefont{Phua}},
  \bibinfo{journal}{Math. Prog.} \textbf{\bibinfo{volume}{14}},
  \bibinfo{pages}{149} (\bibinfo{year}{1978}).

\bibitem[{\citenamefont{Schnabel}(1978)}]{Schnabel78}
\bibinfo{author}{\bibfnamefont{R.~B.} \bibnamefont{Schnabel}},
  \bibinfo{journal}{Math. Programming} \textbf{\bibinfo{volume}{15}},
  \bibinfo{pages}{247} (\bibinfo{year}{1978}).

\bibitem[{\citenamefont{Mart\'inez and Ochi}(1982)}]{Martinez82}
\bibinfo{author}{\bibfnamefont{J.~M.} \bibnamefont{Mart\'inez}}
  \bibnamefont{and} \bibinfo{author}{\bibfnamefont{L.}~\bibnamefont{Ochi}},
  \bibinfo{journal}{Matematica Aplicada e Computational}
  \textbf{\bibinfo{volume}{1}}, \bibinfo{pages}{135} (\bibinfo{year}{1982}).

\bibitem[{\citenamefont{Wolfe}(1959)}]{Wolfe59}
\bibinfo{author}{\bibfnamefont{P.}~\bibnamefont{Wolfe}},
  \bibinfo{journal}{Comm. ACM} \textbf{\bibinfo{volume}{12}},
  \bibinfo{pages}{12} (\bibinfo{year}{1959}).

\bibitem[{\citenamefont{Ortega and Rheinboldt}(1970)}]{OrtegaRheinboldt70}
\bibinfo{author}{\bibfnamefont{J.~M.} \bibnamefont{Ortega}} \bibnamefont{and}
  \bibinfo{author}{\bibfnamefont{W.~C.} \bibnamefont{Rheinboldt}},
  \emph{\bibinfo{title}{Iterative Solution of Nonlinear Equations in Several
  Variables}} (\bibinfo{publisher}{Academic Press}, \bibinfo{address}{New
  York}, \bibinfo{year}{1970}).

\bibitem[{\citenamefont{Gragg and Stewart}(1976)}]{GraggStewart76}
\bibinfo{author}{\bibfnamefont{W.}~\bibnamefont{Gragg}} \bibnamefont{and}
  \bibinfo{author}{\bibfnamefont{G.}~\bibnamefont{Stewart}},
  \bibinfo{journal}{SIAM J. Numer. Anal.} \textbf{\bibinfo{volume}{13}},
  \bibinfo{pages}{889} (\bibinfo{year}{1976}).

\bibitem[{\citenamefont{Mart\'inez}(1979)}]{Martinez79}
\bibinfo{author}{\bibfnamefont{J.~M.} \bibnamefont{Mart\'inez}},
  \bibinfo{journal}{BIT} \textbf{\bibinfo{volume}{19}}, \bibinfo{pages}{236}
  (\bibinfo{year}{1979}).

\bibitem[{\citenamefont{Schnabel}(1983)}]{Schnabel83}
\bibinfo{author}{\bibfnamefont{R.~B.} \bibnamefont{Schnabel}},
  \bibinfo{type}{Tech. Rep.} \bibinfo{number}{CU-CS-247-83},
  \bibinfo{institution}{University of Colorado, Boulder}
  (\bibinfo{year}{1983}).

\bibitem[{\citenamefont{Ford and Moghrabi}(1997)}]{FordMoghrabi97}
\bibinfo{author}{\bibfnamefont{J.}~\bibnamefont{Ford}} \bibnamefont{and}
  \bibinfo{author}{\bibfnamefont{I.}~\bibnamefont{Moghrabi}},
  \bibinfo{journal}{J. Comp. Appl. Math.} \textbf{\bibinfo{volume}{82}},
  \bibinfo{pages}{105} (\bibinfo{year}{1997}).

\bibitem[{\citenamefont{Moreau}(1965)}]{Moreau65}
\bibinfo{author}{\bibfnamefont{J.~J.} \bibnamefont{Moreau}},
  \bibinfo{journal}{Bull. de la Soc. math. de France}
  \textbf{\bibinfo{volume}{93}}, \bibinfo{pages}{273} (\bibinfo{year}{1965}).

\bibitem[{\citenamefont{Rockafellar and Wets}(1998)}]{VA}
\bibinfo{author}{\bibfnamefont{R.~T.} \bibnamefont{Rockafellar}}
  \bibnamefont{and} \bibinfo{author}{\bibfnamefont{R.~J.} \bibnamefont{Wets}},
  \emph{\bibinfo{title}{Variational Analysis}} (\bibinfo{publisher}{Springer},
  \bibinfo{address}{Berlin}, \bibinfo{year}{1998}).

\bibitem[{\citenamefont{Gomes-Ruggiero and Mart\'inez}(1992)}]{Martinez92}
\bibinfo{author}{\bibfnamefont{M.~A.} \bibnamefont{Gomes-Ruggiero}}
  \bibnamefont{and} \bibinfo{author}{\bibfnamefont{J.~M.}
  \bibnamefont{Mart\'inez}}, \bibinfo{journal}{Math. Modelling Numer. Anal.}
  \textbf{\bibinfo{volume}{26}}, \bibinfo{pages}{309} (\bibinfo{year}{1992}).

\bibitem[{\citenamefont{Gomes-Ruggiero
  et~al.}(1992)\citenamefont{Gomes-Ruggiero, Mart\'inez, and
  Moretti}}]{Martinez92b}
\bibinfo{author}{\bibfnamefont{M.~A.} \bibnamefont{Gomes-Ruggiero}},
  \bibinfo{author}{\bibfnamefont{J.~M.} \bibnamefont{Mart\'inez}},
  \bibnamefont{and} \bibinfo{author}{\bibfnamefont{A.~C.}
  \bibnamefont{Moretti}}, \bibinfo{journal}{SIAM J. Sci. Statist. Comput.}
  \textbf{\bibinfo{volume}{13}}, \bibinfo{pages}{459} (\bibinfo{year}{1992}).

\bibitem[{\citenamefont{Mifflin}(1975)}]{Mifflin75}
\bibinfo{author}{\bibfnamefont{R.}~\bibnamefont{Mifflin}},
  \bibinfo{journal}{Math. Program.} \textbf{\bibinfo{volume}{9}},
  \bibinfo{pages}{100} (\bibinfo{year}{1975}).

\bibitem[{\citenamefont{Luke et~al.}(2002)\citenamefont{Luke, Burke, and
  Lyon}}]{Luke02a}
\bibinfo{author}{\bibfnamefont{D.~R.} \bibnamefont{Luke}},
  \bibinfo{author}{\bibfnamefont{J.~V.} \bibnamefont{Burke}}, \bibnamefont{and}
  \bibinfo{author}{\bibfnamefont{R.~G.} \bibnamefont{Lyon}},
  \bibinfo{journal}{SIAM Rev.} \textbf{\bibinfo{volume}{44}},
  \bibinfo{pages}{169} (\bibinfo{year}{2002}).

\bibitem[{\citenamefont{Schnabel and Frank}(21)}]{Schnabel84}
\bibinfo{author}{\bibfnamefont{R.~B.} \bibnamefont{Schnabel}} \bibnamefont{and}
  \bibinfo{author}{\bibfnamefont{P.~D.} \bibnamefont{Frank}},
  \bibinfo{journal}{SIAM J. Numer. Anal.} \textbf{\bibinfo{volume}{21}},
  \bibinfo{pages}{815} (\bibinfo{year}{21}).

\bibitem[{\citenamefont{Schnabel and Frank}(1986)}]{Schnabel86}
\bibinfo{author}{\bibfnamefont{R.~B.} \bibnamefont{Schnabel}} \bibnamefont{and}
  \bibinfo{author}{\bibfnamefont{P.~D.} \bibnamefont{Frank}},
  \bibinfo{type}{Tech. Rep.} \bibinfo{number}{CU-CS-334-86},
  \bibinfo{institution}{University of Colorado, Boulder}
  (\bibinfo{year}{1986}).

\bibitem[{\citenamefont{Borwein and Lewis}(2006)}]{BorLew06}
\bibinfo{author}{\bibfnamefont{J.~M.} \bibnamefont{Borwein}} \bibnamefont{and}
  \bibinfo{author}{\bibfnamefont{A.~S.} \bibnamefont{Lewis}},
  \emph{\bibinfo{title}{Convex analysis and nonlinear optimization : theory and
  examples}} (\bibinfo{publisher}{Springer Verlag}, \bibinfo{address}{New
  York}, \bibinfo{year}{2006}), \bibinfo{edition}{2nd} ed.

\bibitem[{\citenamefont{Browder}(1965)}]{Browder65}
\bibinfo{author}{\bibfnamefont{F.~E.} \bibnamefont{Browder}},
  \bibinfo{journal}{Proc. Nat. Acad. Sci. U.S.A.}
  \textbf{\bibinfo{volume}{54}}, \bibinfo{pages}{1041} (\bibinfo{year}{1965}).

\bibitem[{\citenamefont{Kirk}(1965)}]{Kirk65}
\bibinfo{author}{\bibfnamefont{W.~A.} \bibnamefont{Kirk}},
  \bibinfo{journal}{Amer. Math. Monthly} \textbf{\bibinfo{volume}{72}},
  \bibinfo{pages}{1004} (\bibinfo{year}{1965}).

\bibitem[{\citenamefont{Pratt}(1952)}]{Pratt}
\bibinfo{author}{\bibfnamefont{G.}~\bibnamefont{Pratt}},
  \bibinfo{journal}{Phys. Rev.} \textbf{\bibinfo{volume}{88}},
  \bibinfo{pages}{1217} (\bibinfo{year}{1952}).

\bibitem[{\citenamefont{Prodan and P.}(2003)}]{Prodan03}
\bibinfo{author}{\bibfnamefont{E.}~\bibnamefont{Prodan}} \bibnamefont{and}
  \bibinfo{author}{\bibfnamefont{N.}~\bibnamefont{P.}}, \bibinfo{journal}{J.
  Stat. Phys.} \textbf{\bibinfo{volume}{111}}, \bibinfo{pages}{967}
  (\bibinfo{year}{2003}).

\bibitem[{\citenamefont{Levenberg}(1944)}]{Levenberg}
\bibinfo{author}{\bibfnamefont{K.}~\bibnamefont{Levenberg}},
  \bibinfo{journal}{The Quarterly of Applied Mathematics}
  \textbf{\bibinfo{volume}{2}}, \bibinfo{pages}{164} (\bibinfo{year}{1944}).

\bibitem[{\citenamefont{Marquardt}(1963)}]{Marquardt}
\bibinfo{author}{\bibfnamefont{D.}~\bibnamefont{Marquardt}},
  \bibinfo{journal}{SIAM J. Appl. Math.} \textbf{\bibinfo{volume}{11}},
  \bibinfo{pages}{431} (\bibinfo{year}{1963}).

\bibitem[{\citenamefont{Tihonov}(1963{\natexlab{a}})}]{Tikhonov1}
\bibinfo{author}{\bibfnamefont{A.~N.} \bibnamefont{Tihonov}},
  \bibinfo{journal}{Dokl. Akad. Nauk SSSR} \textbf{\bibinfo{volume}{151}},
  \bibinfo{pages}{501} (\bibinfo{year}{1963}{\natexlab{a}}).

\bibitem[{\citenamefont{Tihonov}(1963{\natexlab{b}})}]{Tikhonov2}
\bibinfo{author}{\bibfnamefont{A.~N.} \bibnamefont{Tihonov}},
  \bibinfo{journal}{Dokl. Akad. Nauk SSSR} \textbf{\bibinfo{volume}{153}},
  \bibinfo{pages}{49} (\bibinfo{year}{1963}{\natexlab{b}}).

\bibitem[{\citenamefont{Hansen}(1998)}]{Hansen98}
\bibinfo{author}{\bibfnamefont{P.~C.} \bibnamefont{Hansen}},
  \emph{\bibinfo{title}{Rank-Deficient and Discrete Ill-Posed Problems}}
  (\bibinfo{publisher}{SIAM}, \bibinfo{year}{1998}).

\bibitem[{\citenamefont{Hoerl and Kennard}(1970)}]{Hoerl}
\bibinfo{author}{\bibfnamefont{A.~E.} \bibnamefont{Hoerl}} \bibnamefont{and}
  \bibinfo{author}{\bibfnamefont{R.~W.} \bibnamefont{Kennard}},
  \bibinfo{journal}{Technometrics} \textbf{\bibinfo{volume}{12}},
  \bibinfo{pages}{55} (\bibinfo{year}{1970}).

\bibitem[{\citenamefont{Raczkowski et~al.}(2001)\citenamefont{Raczkowski,
  Canning, and Wang}}]{Dielectric1}
\bibinfo{author}{\bibfnamefont{D.}~\bibnamefont{Raczkowski}},
  \bibinfo{author}{\bibfnamefont{A.}~\bibnamefont{Canning}}, \bibnamefont{and}
  \bibinfo{author}{\bibfnamefont{L.~W.}~\bibnamefont{Wang}},
  \bibinfo{journal}{Phys. Rev. B} \textbf{\bibinfo{volume}{64}},
  \bibinfo{pages}{121101(R)} (\bibinfo{year}{2001}).

\bibitem[{\citenamefont{Ho et~al.}(1982)\citenamefont{Ho, J., and
  Joanopoulos}}]{Dielectric2}
\bibinfo{author}{\bibfnamefont{K.~M.} \bibnamefont{Ho}},
  \bibinfo{author}{\bibfnamefont{J.}~\bibnamefont{Ihm}}, \bibnamefont{and}
  \bibinfo{author}{\bibfnamefont{J.~D.} \bibnamefont{Joannopoulos}},
  \bibinfo{journal}{Phys. Rev. B} \textbf{\bibinfo{volume}{25}},
  \bibinfo{pages}{4260} (\bibinfo{year}{1982}).

\bibitem[{\citenamefont{Perdew et~al.}(1996)\citenamefont{Perdew, Burke, and
  Ernzerhof}}]{PBE}
\bibinfo{author}{\bibfnamefont{J.~P.}~\bibnamefont{Perdew}},
  \bibinfo{author}{\bibfnamefont{K.}~\bibnamefont{Burke}}, \bibnamefont{and}
  \bibinfo{author}{\bibfnamefont{M.}~\bibnamefont{Ernzerhof}},
  \bibinfo{journal}{Phys. Rev. Lett.} \textbf{\bibinfo{volume}{77}},
  \bibinfo{pages}{3865} (\bibinfo{year}{1996}).

\bibitem[{\citenamefont{Luke}((to appear))}]{Luke07}
\bibinfo{author}{\bibfnamefont{D.~R.} \bibnamefont{Luke}},
  \bibinfo{journal}{SIAM J. Optim.}  (\bibinfo{year}{(to appear)}).

\end{thebibliography}

\end{document}